\documentclass[twocolumn]{aastex7}
\usepackage{caption}
\usepackage{subfigure}
\usepackage{graphicx}
\usepackage{amsmath}
\usepackage{hyperref}

\begin{document}

\title{\Large{Rossby number regime, convection suppression, \\[.2cm] and dynamo-generated magnetism in inflated hot Jupiters}}

\author[orcid=0000-0002-0849-2033,sname='Albert Elias-López']{Albert Elias-L\'opez}
\affiliation{Institut de Ciències de l'Espai (ICE-CSIC), Barcelona, Spain}
\affiliation{Institut d’Estudis Espacials de Catalunya (IEEC), Barcelona, Spain}
\affiliation{Center for Computational Astrophysics, Flatiron Institute, New York, USA}
\email[show]{albert.elias@csic.es}  

\author[orcid=0000-0002-8171-8596,sname='Matteo Cantiello']{Matteo Cantiello} 
\affiliation{Center for Computational Astrophysics, Flatiron Institute, New York, USA}
\affiliation{Department of Astrophysical Sciences, Princeton University, New Jersey, USA}
\email{mcantiello@flatironinstitute.org}

\author[orcid=0000-0001-7795-6850,sname='Daniele Viganò']{Daniele Viganò}
\affiliation{Institut de Ciències de I’Espai (ICE-CSIC), Barcelona, Spain}
\affiliation{Institut d’Estudis Espacials de Catalunya (IEEC), Barcelona, Spain}
\affiliation{Institute of Applied Computing \& Community Code (IAC3), Palma, Spain}
\email{daniele.vigano@csic.es}

\author[orcid=0000-0001-9268-4849,sname='Fabio Del Sordo']{Fabio Del Sordo}
\affiliation{Scuola Normale Superiore, Piazza dei Cavalieri, Pisa, Italy}
\affiliation{INAF, Osservatorio Astrofisico di Catania, Catania, Italy}
\email{fabio.delsordo@sns.it}

\author[orcid=0000-0002-5820-2532,sname='Simranpreet Kaur']{Simranpreet Kaur}
\affiliation{Institut de Ciències de I’Espai (ICE-CSIC), Barcelona, Spain}
\affiliation{Institut d’Estudis Espacials de Catalunya (IEEC), Barcelona, Spain}
\email{simranpreet.k@csic.es}  

\author[orcid=0000-0002-3070-1647,sname='Clàudia Soriano Guerrer']{Clàudia Soriano-Guerrero}
\affiliation{Institut de Ciències de I’Espai (ICE-CSIC), Barcelona, Spain}
\affiliation{Institut d’Estudis Espacials de Catalunya (IEEC), Barcelona, Spain}
\email{claudia.soriano@csic.es} 

\begin{abstract}
    Hot Jupiters (HJs) are commonly thought to host the strongest dynamo-generated magnetic fields among exoplanets, up to one order of magnitude larger than Jupiter. Thus, they have often been regarded as the most promising exoplanets to display magnetic star-planet interaction signals and magnetically-driven coherent radio emission, which unfortunately remains elusive, despite many diversified observational campaigns. In this work, we investigate the evolution of the internal convection and dynamo properties of HJs via one-dimensional models. We explore the dependency on orbital distance, planetary and stellar masses, and types of heat injection. We employ one-dimensional evolutionary models to obtain internal convective structures. Specifically, we obtain the Rossby number $\mathrm{Ro}$ as a function of planetary depth and orbital period, after showing that tidal synchronization is likely valid for all HJs. When the heat is applied uniformly, the convective layers of almost all HJs remain in the fast rotator regime, $\mathrm{Ro} \lesssim 0.1$, except possibly the most massive planets with large orbital distances (but still tidally locked). We recover magnetic field strengths for inflated HJs by applying well-known scaling laws for fast rotators. When strong heat sources are applied mostly in the outer envelope and outside the dynamo region, as realistic Ohmic models predict, convection in the dynamo region often breaks down. Consequently, the heat flux and the derived surface magnetic fields can be greatly reduced to or below Jovian values, contrary to what is commonly assumed, thus negatively affecting estimates for coherent radio emission, and possibly explaining the failure in detecting it so far.
\end{abstract}


\section{Introduction}
\label{Sec: Introduction}

Hot Jupiters (HJs) are gaseous giant planets on close-in orbits, characterized by orbital periods of just a few days and separations  $\lesssim 0.1$ AU \citep{Fortneyetal2021}. Due to the proximity to their host stars, they are highly irradiated and tidally locked, with equilibrium temperatures up to 3000 K. This extreme external heating creates strong thermal gradients between their day and night sides, driving strong atmospheric flows and equatorial zonal jets that shape their atmospheric dynamics (e.g., \citealt{Showman&Guillot2002, Showmanetal2009, Heng&Showman2015}). Currently, several hundred HJs have been confirmed, enabling extensive population studies. One of the most intriguing findings is the so-called HJ radius anomaly (see \cite{Thorngren2024} for a review), which remains one of the longest-standing open questions in exoplanetary science: observations show that HJs systematically exhibit inflated radii, significantly larger (up to twice the Jovian radius) than those predicted by standard evolutionary models. There is a clear trend between the degree of radius inflation and stellar irradiation, once the planetary equilibrium temperature exceeds $T_{\rm eq}\gtrsim 1000$ K \citep{Weissetal2013}, a fundamental hint to the underlying physical mechanism. However, the irradiation alone can contribute to radius inflation only up to approximately $\sim 1.3~R_J$ at Gyr ages, depending also on the mass and the amount of heavy elements \citep{Guillotetal96, Guillot&Showman2002, Arras&Bildsten2006, Burrowsetal2007, Fortneyetal2007}. Several mechanisms have been proposed to delay the shrinking and cooling \citep{Spiegel&Burrows2013, Thorngren2024}: enhanced opacities \citep{Burrowsetal2007}, inhibition of large-scale convection \citep{Chabrier&Baraffe2007}, hydrodynamic effects leading to dissipation in deep layers \citep{Showman&Guillot2002, Showmanetal2009, Li&Goodman2010, Youdin&Mitchell2010}, and Ohmic dissipation of atmospherically induced magnetic field. Among these proposed mechanisms, the latter has received considerable attention, with several quantitative studies (e.g., \citealt{Batygin&Stevenson2010, Batyginetal2011, Pernaetal2010b, Wu&Lithwick2013, Ginzburg&Sari2016, Knierimetal2022, SorianoGuerreroetal2023, SorianoGuerreroetal2025, Viganoetal2025}). But multiple mechanisms may operate simultaneously \citep{Sarkisetal2021}.

One of the most elusive characteristics of exoplanets is their magnetism. Observational estimates of magnetic fields in some HJs, with considerable uncertainties, have been proposed so far in very few works, via star-planet interaction (SPI) models, used to interpret the observed trends of X-ray luminosities vs presence of short-orbit planets \citep{Scharfetal2010}, or the modulation of Ca II K lines with HJ orbital periods \citep{Cauleyetal2019}. 
Self-sustained dynamo simulations for the planetary scenario have been used to derive scaling laws that relate various dimensionless parameters \citep{Christensen&Aubert2006, Yadavetal2013}, which have been validated by the comparison with the observed magnetic fields on Earth, Jupiter, and in fast-rotating low-mass stars \citep{Christensenetal2009, Reiners&Christensen2010}. These scaling laws state that, in the fast rotator regime, convective heat flux determines the magnetic field strength. When applied in the HJ context, heat-flux scaling laws generally found an increase by up to $\sim 1$ order of magnitude of the inferred magnetic fields compared to non-inflated gas giants, if the extra heat needed to explain the inflation is assumed to take the role of the heat convective flux in the scaling laws \citep{Yadav&Thorngren2017, Kilmetisetal2024}. Note that, on the contrary, slowly rotating cool stars observationally show a correlation of the magnetic activity indicators with the rotation rate \citep{Reinersetal2014}. The transition from a rotation-dominated (slow rotators) to a convective-heat-flux-dominated (fast rotators) dynamo strength is seen to be fairly sharply defined by the Rossby number $\mathrm{Ro} \sim 0.12$. In this sense, planetary scaling laws applicable to slow rotators assumed a proportionality with the rotation rate rather than the heat flux (e.g., \citealt{Sanchezlavega2004, Stevens2005}). A primary aim of this work is to shed light on which regime applies to HJs, considering the observed range of their relevant parameters: planetary mass, stellar type, and star-planet separation.

Moreover, since virtually all the existing scaling laws predict a correlation between magnetic field intensity and planetary mass, giant exoplanets, and HJs in particular, have been historically considered the main candidates for the still elusive quest of Jovian-like coherent radio emission (e.g., \citealt{deGasperinetal2020, Narangetal2024} and references within), in the form of electron cyclotron maser \citep{Dulk1985}. The latter has been detected from all magnetized planets in the solar system \citep{Zarka1998}, but can be detected from the ground only if the cyclotron frequency (ECM), $\nu_c \simeq 2.8~ B$ [G] MHz, is larger than the $\sim 10$ MHz ionospheric cut-off. This implies that, to be detected by ground-based radio interferometers, the magnetic field needs to be at least as intense as the Jovian one. In this sense, the second main aim of this work is to explore the trends of the estimates of magnetic field intensity at the dynamo region and the planetary surface, considering their Rossby regime and applying the appropriate scaling law at a given age, with the internal structure and (convective) luminosity given by long-term evolutionary models. Understanding the magneto-thermal dependence on orbital distance and age can help guide future observations toward the most promising targets.

\begin{figure*}[ht!]\centering
\includegraphics[width=.4\textwidth]{ 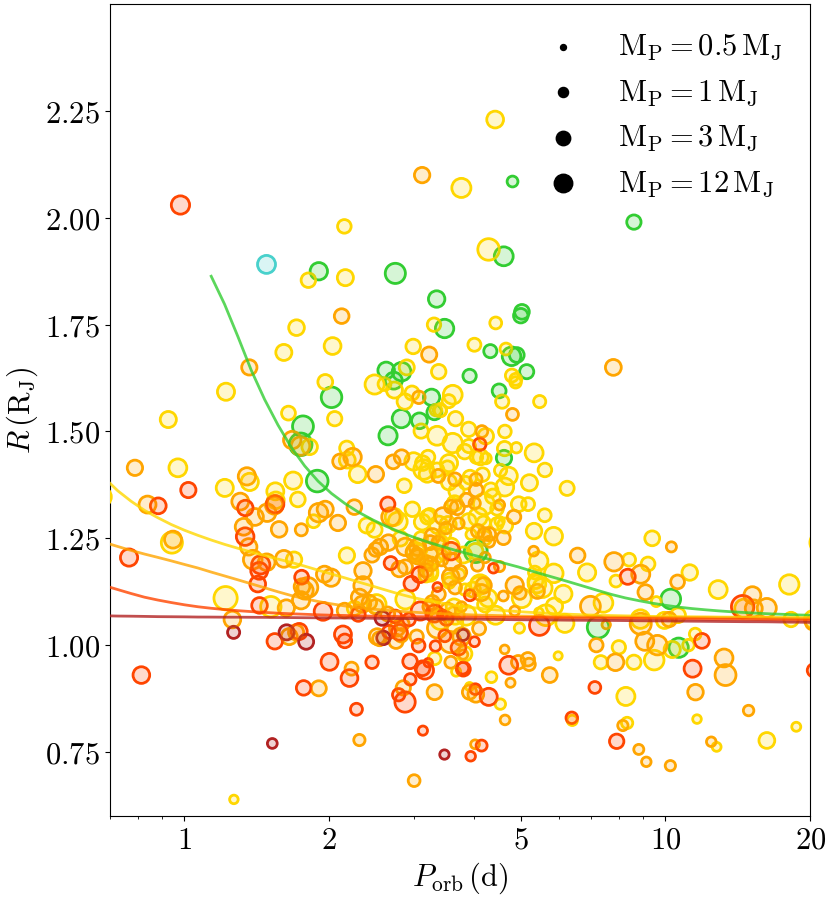}\includegraphics[width=.595\textwidth]{ 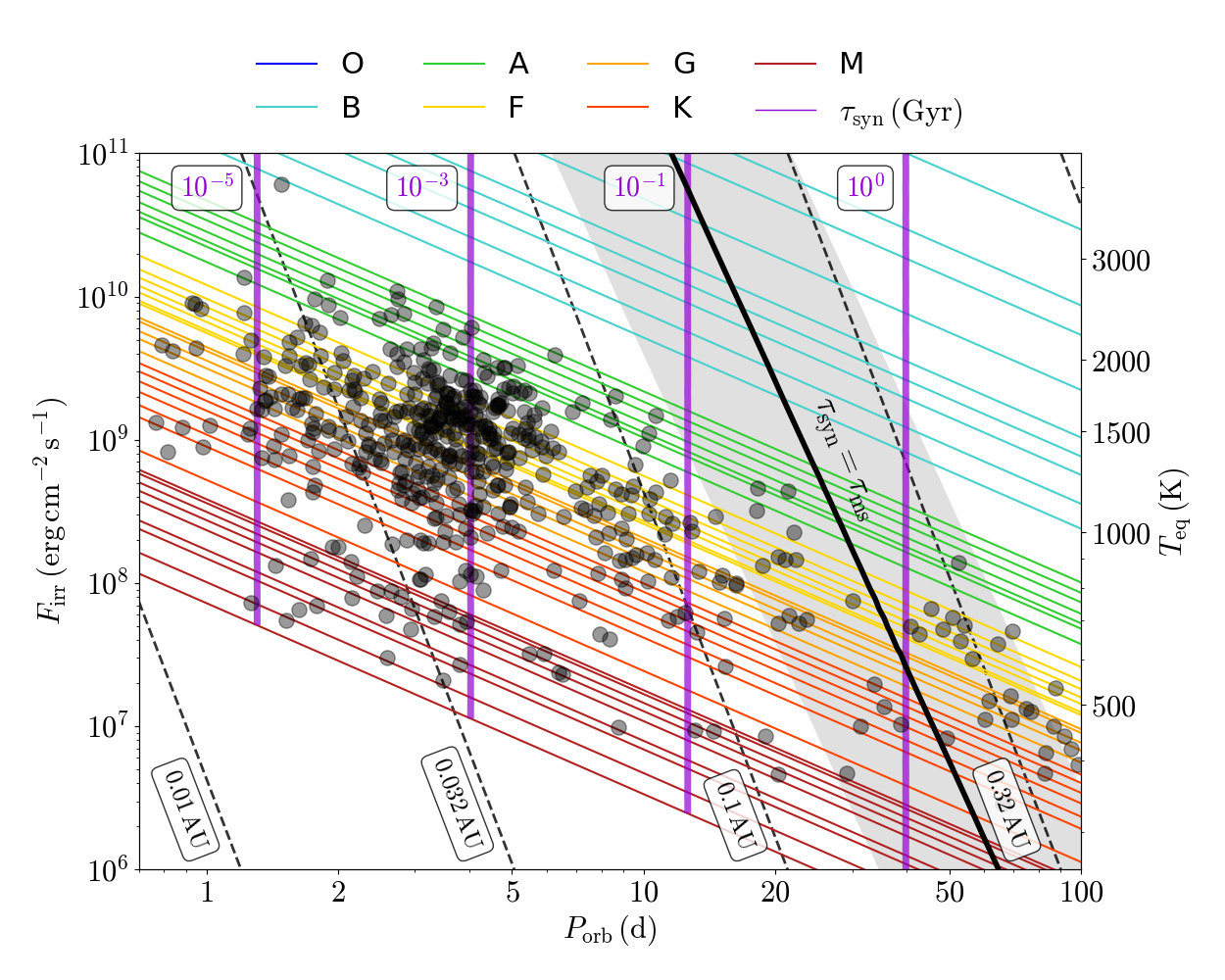}
\caption{Population properties for the confirmed gas giant exoplanet population with mass M$_\text{P}$ $>$ 0.2 M$_\text{J}$, considering the host star effective temperature and radius to obtain the irradiation flux $F_{\rm irr}$, and the corresponding equilibrium temperature $T_{\rm eq}$. Data is extracted from the NASA exoplanet archive\footnote{\url{https://exoplanetarchive.ipac.caltech.edu/}} \citep{Christiansenetal2025}. Colors indicate the main-sequence stellar type of the host star. Errors are not shown, for the sake of visibility. {\em Left:} planetary radius as a function of orbital period. Marker size denotes the planetary mass. {\rm Right:} $F_{\rm irr}$ as a function of planetary orbital period for the same exoplanetary systems. The dashed black lines mark constant separation $a$. The purple vertical lines correspond to different synchronization timescales, $\tau_{\rm syn}$ in Gyr, obtained with Eq.~\ref{Eq: Tau sync}, assuming $M_{\rm P} = 1~\mathrm{M_J}$, $Q'_P$ = 5 $\cdot$ 10$^5$, $\alpha=2/5$, and $\omega_i=2 \pi/5 \, \mathrm{h}^{-1}$ (see text). The solid black line indicates where the main-sequence lifetime equals the synchronization timescale.}
\label{Fig: Hot Jupiter fluxes and radii}
\end{figure*}

This work is organized as follows. In Sect.~\ref{Sec: Irradiation and tidal synchronization} we summarize the values of irradiation and tidal locking timescales, based on the HJ population. Sect.~\ref{Sec: Planetary evolution} introduces the evolutionary models and the amount of additional internal heat as a function of irradiation, based on the statistical study by \cite{Thorngren&Fortney2018}. We also discuss how the injection and irradiation affect the internal convective layers. In Sect.~\ref{Sec: Rossby number estimation}, we obtain internal profiles of the local Rossby number and discuss how it depends on age, star-planet separation, planetary mass, and stellar type. In Sect.~\ref{Sec: Observational consequences} we estimate the dynamo-generated magnetic fields, and the corresponding values at the planetary surfaces, via the scaling laws mentioned above. We also discuss the related implications for magnetic SPI and ECM radio emission. Finally, we highlight the main conclusions in Sect.~\ref{Sec: Conclusions}.

\section{Irradiation and tidal synchronization}
\label{Sec: Irradiation and tidal synchronization}

The irradiation flux received by a planet in a specific circular orbit of radius $a$ (hereafter, separation) around its host star is:
\begin{equation}
    F_{\rm irr} = \frac{L_{\star}}{4 \pi a^2} = \sigma_{\rm SB} T_{\rm eff}^4 \,\frac{R_{\star}^2}{a^2}\,,
\label{Eq: Hot Jupiter irradiation}
\end{equation}
where $R_\star$, $L_{\star}$ and $T_{\rm eff}$ are the stellar radius, luminosity and effective temperature, respectively. Stellar luminosities are taken from tabulated main-sequence star values\footnote{In this work, we use the tables of \url{https://sites.uni.edu/morgans/astro/course/Notes/section2/spectralmasses.html}} (i.e., obtained from mass-luminosity relations, $L\propto M^a$, where $a$ depends on stellar mass). The resulting irradiation fluxes for HJs typically range from $10^8$ to $10^{10}$ erg cm$^{-2}$ s$^{-1}$, orders of magnitude larger than the fluxes reaching Earth and Jupiter ($1.4\cdot10^6$ and $5\cdot 10^4$ erg cm$^{-2}$ s$^{-1}$, respectively). The corresponding planetary equilibrium temperature is:
\begin{equation}
    T_{\rm eq} = \left( \frac{F_{\rm irr} (1-A_B)}{4\sigma_{\rm SB}}~\right)^{1/4} = (1-A_B)^{1/4} \left( \frac{R_{\star}}{2a}~\right)^{1/2} T_{\rm eff}\,,
\label{Eq: Hot Jupiter equilibrium temperatures}
\end{equation}
where $A_B$ is the planetary albedo (which we consider zero for simplicity) and $\sigma_{\rm SB}$ is the Stefan-Boltzmann constant.

The left panel of Fig.~\ref{Fig: Hot Jupiter fluxes and radii} shows planetary radii, $R_{\rm P}$, as a function of the orbital period $P_{\rm orb}$ for the currently known HJ population with masses $M_P> 0.2~M_{\rm J}$. In the right panel, we also show the irradiation flux $F_{\rm irr}$ and equilibrium temperature $T_{\rm eq}$, as a function of $P_{\rm orb}$, for the same HJ population. Colors correspond to the spectral type of the host star. The bulk of known HJs orbit F, G, and K-type stars, with orbital distances lower than 0.1 AU, i.e. periods $P_{\rm orb}\lesssim 10$ days. The lines in the left panel correspond to the radii of standard planetary evolutionary models with $1~\mathrm{M_J}$ masses, at 5 Gyr, considering only irradiation and no internal extra heat injection (see Sect.~\ref{Sec: Hydrostatic model}) around different stars (see Table~\ref{Tab: HJ common stellar types} for details). Note that most planets are above these lines, showing signs of inflation and the need for a heat injection mechanism. The HJ population below these lines is mostly small planets compatible with high metallicities or large rocky cores \citep{Laughlin&Crismani2011}, possibly related to early severe mass losses \citep{Sestovicetal18, Lazovik2023}, which are factors that we do not consider for our evolutionary models, since we aim at studying the average properties of HJs, i.e. with inflated radii. Note that, as mentioned before, the clear correlation is between $R_{\rm P}$ and $F_{\rm irr}$ (or $T_{\rm eq}$). However, as we use $P_{\rm orb}$ to define the Rossby number and dynamo regime (see Sect.~\ref{Sec: Rossby number estimation}), we choose it as a key variable, instead of the stellar irradiation $F_{\rm irr}$. Due to the degeneracy between stellar mass and $P_{\rm orb}$ in defining $F_{\rm irr}$, the correlation $R_P$ vs $P_{\rm orb}$ (left panel) is essentially lost, in agreement with e.g. \cite{Weissetal2013}.

Observed HJs are commonly seen to sit on low-eccentricity orbits \citep{Zink&Howard2023}, and assumed to be tidally locked, meaning that their rotational and orbital periods $P_{\rm orb}$ are identical. Secondary transits
finding azimuthal displacements of the hottest points compared to the substellar points to the East (e.g. \citealt{Knutsonetal2007}) or West (e.g. \citealt{Dangetal2018}), are usually interpreted in terms of atmospheric circulation effects, rather than a deviation from the tidal locking regime. Here, we revisit with some detail the quantitative assessment of the timescale $\tau_{\rm syn}$ required to achieve tidal synchronization, by calculating the rate of change of the planetary rotation $\omega$ under the assumption of circular orbits and zero obliquity:
\begin{equation}
\begin{aligned}
  \frac{d\omega}{dt} = \frac{9}{4} \frac{1}{\alpha Q_P'} \frac{G M_{\rm P}}{R_{\rm P}^3} \Bigg(\frac{M_*}{M_{\rm P}}\Bigg)^2 \Bigg(\frac{R_{\rm P}}{a}\Bigg)^6 \\ \quad \text{where} \quad  \alpha = \frac{I}{M_{\rm P} R_{\rm P}^2} \quad , \quad Q_{\rm P}' = \frac{3 Q_{\rm P}}{2k_{\rm 2,P}}\,,
\end{aligned}
\label{Eq: Rotation period tidal evolution}
\end{equation}
$M_{\rm P}$ and $R_{\rm P}$ are the planetary mass and radius, respectively, $I$ is the planetary moment of inertia, $Q_{\rm P}$ is the dissipation factor, $k_{\rm 2,P}$ is the Love number of the planet and $G$ is the gravitational constant \citep[e.g.][]{Goldreich&Soter1966, Murray&Dermott1999, Griessmeieretal2007}. The synchronization timescale is then given by:
\begin{equation}
\begin{aligned}
\tau_{\rm syn}\approx\frac{\Delta \omega}{d\omega / dt} = \frac{4}{9} \frac{d^6 I Q_P'\Delta \omega}{G M_*^2 R_{\rm P}^5} \simeq \frac{G \alpha}{36 \pi^4} \frac{M_{\rm P} Q_P' P^4 \omega_i}{R_{\rm P}^3}~,
\end{aligned}
\label{Eq: Tau sync}
\end{equation}
where $P$ is the orbital period, and $\Delta \omega = \omega_i - \omega_f \sim \omega_i$, since $\omega_i \gg \omega_f$, the difference between initial and final rotation. 
On the right panel of Fig.~\ref{Fig: Hot Jupiter fluxes and radii} we show in purple different lines of constant $\tau_{\rm syn}$ for a $M_{\rm P} = 1~ \mathrm{M_J}$ planet, assuming $\alpha=2/5$ (homogeneous sphere), $R_{\rm P}$ = 1.5 $\mathrm{R_J}$, $Q'_P$ = 5 $\cdot$ 10$^5$ as in \cite{Griessmeieretal2007}, and an initial spin period of 5 hours typical of known fast-rotating sub-stellar objects, $\omega_i=(2 \pi/5$ h), and conservatively larger than the breakup values which can be used as initial condition \citep{Batygin2018}. Most of the parameters adopted in the above expression evolve significantly over time, particularly at early stages ($t \lesssim 10$ Myr) when planetary contraction proceeds rapidly, and orbital migration as well as interactions with the protoplanetary disk are important. For this reason, we adopted reference values that yield systematically conservative estimates,  overestimating $\tau_{\rm syn}$ by factors of a few. Specifically, realistic adjustments would include considering a radially dependent density profile (moment of inertia factor $\alpha<2/5$), larger radii of very young gas giants depending on their formation entropy \citep[e.g.,][]{Fortneyetal2007}, and shorter initial spin periods, which can be as brief as about 2 hours for newborn brown dwarfs and low-mass stars.

We show in Fig.~\ref{Fig: Hot Jupiter fluxes and radii} the location where $\tau_{\rm syn}$ equals the main sequence lifetime of the host star as a black line. We calculated the duration of the main sequence as $\tau_{\rm ms} = 10~\mathrm{Gyr}~(M_\star/M_\odot)^{-2/5}$ \citep[chap. 13]{Carroll&Ostlie2017}. Planets lying to the left of this line are expected to be tidally locked, or to reach that state within the stellar lifetime. To evaluate how planetary mass affects $\tau_{\rm syn}$, we observe that for a 0.2 $\mathrm{M_J}$, 2 $R_J$ planet, $\tau_{\rm syn}$ becomes a factor of 10 smaller. On the other hand, a massive $M_{\rm P}$ = 10 $\mathrm{M_J}$ takes about one order of magnitude longer to synchronize. This range of values is shown by the gray shading around the $\tau_{\rm syn} = \tau_{\rm ms}$ (and each $\tau_{\rm syn}$ line has a similar uncertainty). From Fig.~\ref{Fig: Hot Jupiter fluxes and radii}, it is clear that the HJs population with $T_{\rm eq}>1000$ K should be tidally locked, and only warm Jupiters (having $P_{\rm orb}\gtrsim 20-30$ d) could still preserve their initial faster rotation.

Note that, recently, \cite{Wazny&Menou2025} used global circulation models to show that the presence of a non-negligible magnetic torque in atmospheres permeated by thermally driven winds could lead to substantial spin-orbit asynchronization, particularly in planets with high equilibrium temperatures and strong magnetic fields. These effects are beyond the scope of this paper, and thus, we adopt the assumption of tidal synchronization throughout our analysis.

\section{Planetary evolution}
\label{Sec: Planetary evolution}

\subsection{Cooling model}
\label{Sec: Hydrostatic model}

We use the public code MESA\footnote{\url{https://mesastar.org/}} \citep{Paxtonetal2011, Paxtonetal2013, Paxtonetal2015, Paxtonetal2018, Paxtonetal2019, Jermynetal2023}, specifically its version 24.08.1, to evolve irradiated gas giants, allowing for additional internal heat deposition. The code solves the time-dependent one-dimensional stellar structure equations, using the adapted module for gas giants \citep{Paxtonetal2013}. The set of equations includes mass conservation, hydrostatic equilibrium, energy conservation, and energy transport:
\begin{gather}
    \frac{dm}{dr} = 4 \pi r^2~\rho ~,
    \label{Eq: MESA mass conservation}\\
    \frac{dP}{dm} = - \frac{G m}{4\pi r^4} ~,
    \label{Eq: MESA hydrostatic equilibrium}\\
    \frac{dL}{dm} = - T \frac{ds}{dt} + \epsilon_{\rm irr} + \epsilon_{\rm heat} ~,
    \label{Eq: MESA luminosity}\\
    \frac{dT}{dm} = - \frac{G m T}{4\pi r^4 P} \nabla ~,
    \label{Eq: MESA energy transport}
\end{gather}
where $m$ is the mass enclosed within a radius $r$, $\rho$ is the density, $P$ is the pressure, $G$ the gravitational constant, $s$ the specific entropy, $T$ the temperature, $L$ the internal luminosity, and $\nabla \equiv d \ln T / d \ln P$ is the logarithmic temperature gradient, which is set to the smallest between the adiabatic gradient and the radiative gradient. The right-hand side of the energy equation~(\ref{Eq: MESA luminosity}) includes the specific heating/cooling rates: gravitational contraction, stellar irradiation $\epsilon_{\rm irr}$ \citep{Guillotetal96}, and any internal heat source $\epsilon_{\rm heat}$. The latter has been included in several previous works using MESA, to consider the effect of continuous heat deposition and inflation in HJs \citep{Komacek&Youdin2017, Thorngren&Fortney2018, Komaceketal2020}. The set of equations is closed using the MESA equation of state \citep{Paxtonetal2019}, which, for the gas giant ranges of interest, is essentially the interpolation of the \cite{Saumonetal1995} equation of state for H-He mixtures.

The atmospheric irradiation is implemented in a simplified way, as in previous MESA HJ works \citep{Paxtonetal2013, Komacek&Youdin2017, Komaceketal2020}: the specific energy absorption rate is assumed to be $\epsilon_{\rm irr} = F_{\rm irr} / (4 \Sigma_\star)$, i.e. a uniform deposition of heat through the outermost mass column $m(r) \leq \Sigma_\star$, where $\Sigma_\star$ parametrises how deep is the column which atmospheric absorption occurs, without a more detailed modeling of the atmospheric composition and opacities. We fix it to 200 g cm$^{-2}$, which approximately corresponds to a grey opacity $\kappa=5\times 10^{-3}$~cm$^2$~g$^{-1}$, with a maximum depth of a fraction of a bar. We consider a constant-in-time value of the stellar irradiation $F_{\rm irr}$, thus neglecting the slight increase in stellar luminosity $L_\star(t)$ throughout the main sequence phase. We refer to  \cite{Lopez&Fortney2016, Komaceketal2020} for works including this effect, important especially at the end of the main sequence and beyond it.
As commonly done in similar previous works, we assume here an inert homogeneous rocky core of mass 10 $M_{\oplus}$ and homogeneous density $\rho_c=10$~g~cm$^{-3}$ (which together set the core size $R_{\rm core}$), and a fixed solar composition for the envelope. We don't discuss here the effects of the total metallicity, and the comparison between a diluted or solid core, since they have been shown in detailed cooling studies for HJs (e.g., \citealt{Burrowsetal2007, Thorngrenetal2016}) and Solar planets (e.g., \citealt{Wahletal2017, Yildizetal2024}).

Generally speaking, during the initial stages ($t\lesssim 10$ Myr), the evolution of the planetary radius depends on the assumed initial condition (i.e., the internal entropy at formation). At later times, instead, models are insensitive to it, and the planet shrinks during its long-term evolution due to its cooling (e.g. \citealt{Burrowsetal2007, Paxtonetal2013}). Therefore, the detailed value of the initial condition is not relevant. To help MESA numerical convergence at early ages for extremely irradiated models, we also use a short relaxation phase at the beginning of evolution, after applying the irradiation.

\begin{table}[t]
\caption{Characteristics of the stars considered throughout the text (specifically the planetary irradiated models in Fig.~\ref{Fig: Hot Jupiter fluxes and radii} and in App.~\ref{App: Stellar type dependence}.}
\begin{tabular}{cccc}
\hline \hline \\[-2.0ex]
Stellar type & $T_{\star}$ [K] & $L_{\star}$ [$\mathrm{L_\odot}$]  & $M_{\star}$ [$\mathrm{M_\odot}$] \\[1.ex] \hline \\[-3ex]
AV3           & 8750  & 12    & 1.86  \\
FV5           & 6700  & 2.4   & 1.25  \\
GV2           & 5800  & 1     & 1.00  \\
KV3           & 4800  & 0.31  & 0.746 \\
MV3           & 3500  & 0.046 & 0.463 \\ \hline \hline \\[-2.0ex]
\end{tabular}
\label{Tab: HJ common stellar types}
\end{table}

We denote the depth at which irradiation reaches $R_{\rm irr}$, and the radiative-convective boundary as $R_{\rm RCB}$. We also define the outer radius of the dynamo region, $R_{\rm dyn}$, by the pressure above which hydrogen becomes metallic, which we take as $P_{\rm dyn}=1$ Mbar \citep{Frenchetal2012}. Note that, strictly speaking, the transition to the metallic hydrogen is not sharp and depends on the temperature, as well \citep{Bonitzetal2024}. However, for simplicity, and given the high theoretical uncertainties about the exact location of the phase transition, we consider a single value of $P_{\rm dyn}$ as a first approximation. In the metallic hydrogen region, the electrical conductivity is $\sigma \sim 10^6$ S m$^{-1}$ \citep{Frenchetal2012, Bonitzetal2024}, orders of magnitude larger than in the molecule-dominated envelope, where the conductivity is dominated by the thermal ionization of alkali metals (e.g., \citealt{Kumaretal2021}). Considering a typical convective velocity $v_{\rm conv}\sim 0.1$ m s$^{-1}$ \citep{Fuentesetal2023}, a Jupiter-like shell thickness $L \sim 5\cdot 10^7$ m, the resulting magnetic Reynolds number in the dynamo region is $\mathrm{Rm} = v_{\rm conv} \, L / \eta = \mu_0 \, \sigma \, v_{\rm conv} \, L \sim 10^6-10^7$ (where $\eta$ is the magnetic diffusivity and $\mu_0$ is the magnetic permeability in vacuum), well above the minimum values commonly thought to be needed for dynamo, $\mathrm{Rm} \gtrsim 50$ \citep{Christensen&Aubert2006}.

Hereafter, we consider planets with 1 d $\lesssim P_{\rm orb} \lesssim$ 30 d orbiting a representative sample of five main-sequence stars, listed in Table~\ref{Tab: HJ common stellar types}. We test planetary masses from 0.5 to 12 $\mathrm{M_J}$, sampling the whole inflated HJ population in period and flux as seen on the right of Fig.~\ref{Fig: Hot Jupiter fluxes and radii}. We restrict to planets $M_P \geq 0.5 M_{\rm J}$ due to the above-mentioned contamination of partially evaporated, high-metallicity planets, as discussed also in e.g. \cite{Sestovicetal18}.

\subsection{Extra heat}
\label{Sec: Extra deposited heat}

Since the radius inflation, which calls for continuous internal heat deposition, correlates with the irradiation flux \citep{Weissetal2013}, a commonly used parameter is the heating efficiency $\epsilon$, defined as the ratio between the total deposited heat rate, $Q_{\rm dep}=\int_M \epsilon_{\rm heat}~{\rm d}m$, and the irradiation flux integrated over the planetary surface:
\begin{equation}
    \epsilon = \frac{Q_{\rm dep}}{\pi R^2 F_{\rm irr}}~.
    \label{Eq: Total deposited heat}
\end{equation}
Values of efficiency of a few percent or less are enough to inflate planets to the observed radii, if the heat is deposited in the convection region (e.g. \citealt{Batygin&Stevenson2010, Batyginetal2011, Wu&Lithwick2013, Komacek&Youdin2017}). A detailed statistical study by \cite{Thorngren&Fortney2018} derived an analytical expression for the amount of deposited heat fraction as a function of the incident flux, $\epsilon(F_{\rm irr})$, that best fits the trend $R_{\rm P}(F_{\rm irr})$ seen in the entire HJ population:

\begin{equation}
    \epsilon = \left( 2.37^{+1.3}_{-0.26} \%~\right) \exp \left[ -\frac{\left(\log_{10} (F_{\rm irr}/F_0) - 0.14^{+0.060}_{-0.069}\right)^2}{2 \cdot \left( 0.37^{+0.038}_{-0.059}~\right)^2}~\right]~,
    \label{Eq: Thorngen and Fortney heat}
\end{equation}
where $F_0=10^9~{\rm erg~ s}^{-1} {\rm cm}^{-2}$. This inferred efficiency shows a strong, non-monotonic dependence on the stellar irradiation: the maximum heating efficiency grows with $T_{\rm eq}$ to reach a maximum of $\sim 2.5 \%$ at $T_{\rm eq} \approx 1500 \,$ K and then decreases, having $\sim 0.2 \%$ at $T_{\rm eq} \approx 2500$~K. Such a non-monotonic trend fits very well with the Ohmic models, for which, as the induced field increases with $T_{\rm eq}$, the Lorentz forces exert an effective larger drag on the thermal winds (e.g., \citealt{Pernaelat2010, Menou2012}). As a consequence, very hot Jupiters might have slower winds, which limit the available energy for the magnetic induction and decrease the overall heating efficiency.

Note that the expression above is purely phenomenological and shows high standard deviations, reflecting the large HJ radius dispersion at a given $T_{\rm eq}$. In a similar work, \cite{Sarkisetal2021} obtain a similar fit with a Gaussian function for the flux, but with the efficiency peaking at $T_{\rm eq} \approx1850 \, K$. Moreover, both \cite{Thorngren&Fortney2018} and \cite{Sarkisetal2021} note that this heating efficiency is degenerate with heavy mass fraction. Despite these uncertainties and caveats, the analytical function above, which we use in this study, offers an easy way to explore the representative internal heating we can infer for a given planet and star.

\begin{figure}[t]
\centering
\includegraphics[width=\hsize]{ 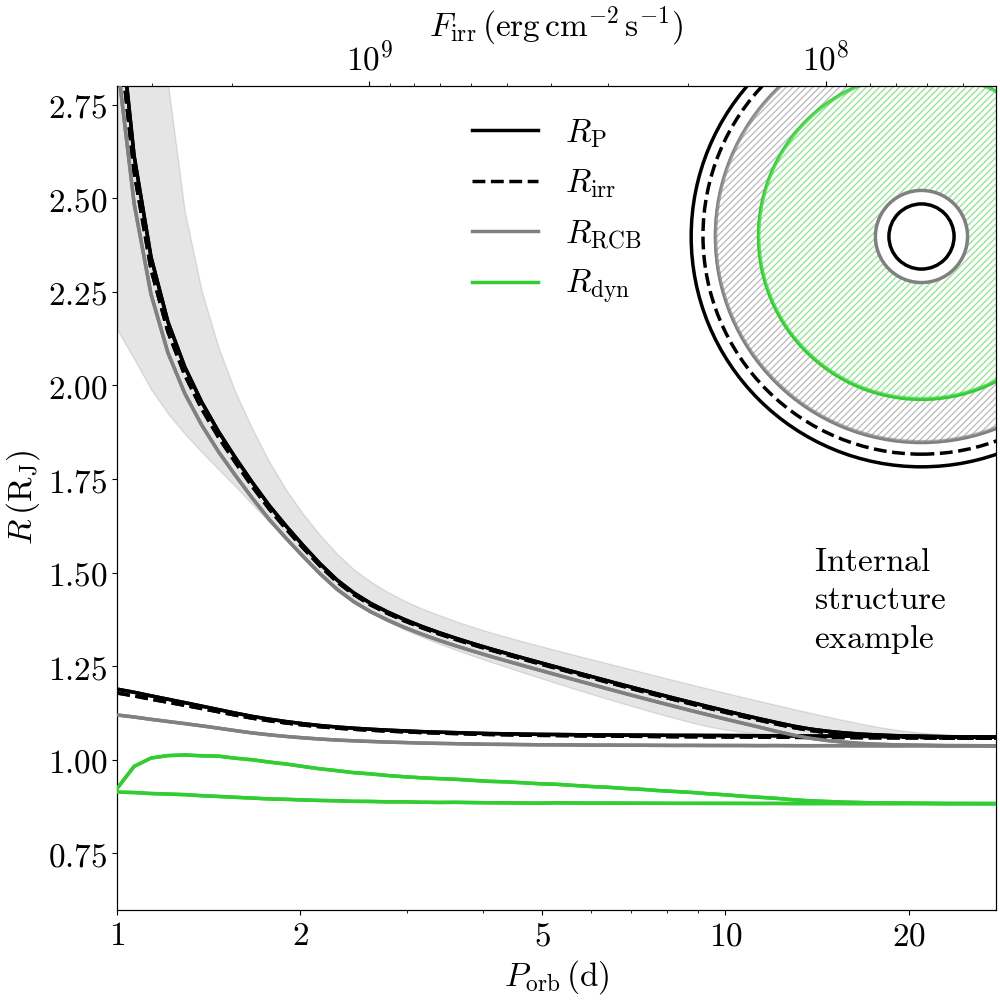}
\caption{Dynamo outer surface radius $R_{\rm dyn}$ (green), RCB radius $R_{\rm rcb}$ (gray), bottom of the irradiated region $R_{\rm irr}$ (dashed), and planetary radius $R_{\rm P}$ (solid black), for a 1 $\mathrm{M_J}$ planet orbiting a 1 $\mathrm{M_\odot}$ star, at 5 Gyr, as a function of orbital period, for both heated (upper lines) and non-heated models (lower lines). The gray area shows the outer radius uncertainty for the heated models, accounting for the range of parameters quoted in Eq. (\ref{Eq: Thorngen and Fortney heat}). In the upper right corner, we show a sketch of the relative position, not at scale, of the different radii, including the internal core $R_{\rm core}$ (solid black) and a possible stratified layer close to the core (gray).}
\label{Fig: Radius as a function of orbit}
\end{figure}

The efficiency is not the only parameter to consider: it is also important to specify where the heat is deposited, i.e., the $\epsilon_{\rm heat}(r)$ profile \citep{Ginzburg&Sari2015, Komacek&Youdin2017}. Since we are not assuming any particular mechanism responsible for the extra heat, for simplicity we inject a uniform specific heat rate, $\epsilon_{\rm heat} = Q_{\rm dep}/M_{\rm shell}$ (heat per unit time per unit mass), over a given shell with mass $M_{\rm shell}$, delimited by two radii, $r_{\rm bot}$ and $r_{\rm top}$, for which we consider two simple scenarios:
\begin{enumerate}
    \item Extended deposition in the dynamo region ($\epsilon_{\rm heat, extended}$): $r_{\rm bot} = R_{\rm core}$; $r_{\rm top} = R_{\rm dyn}$.
    \item Deposition in the outer convective region, outside the dynamo region ($\epsilon_{\rm heat, outer}$): $r_{\rm bot} = R_{\rm dyn}$; $r_{\rm top} = R_{\rm irr}$.
\end{enumerate}
Note that if one considers heating up to the irradiated layer ($r_{\rm bot} = R_{\rm core}$; $r_{\rm top} = R_{\rm irr}$), i.e. including part of the stratified, radiative layers, the results are almost indistinguishable results from the first one. As noted in previous works \citep{Komacek&Youdin2017}, heating the radiative region is less effective in terms of inflated radii, because the mass in the radiative region is very small and the heat mostly escapes outward. Moreover, for very high values of irradiation and heating, including the outermost layers, the code can fail to converge more easily.

The second type of heating injection (i.e., above the dynamo radius) is inspired by the fact that most heating mechanisms deposit the bulk of energy in low-density regions. Physical models such as ohmic, tidal, or turbulent heating have different heating distributions \citep{Batygin&Stevenson2010, Batyginetal2011, Wu&Lithwick2013, Ginzburg&Sari2015, Ginzburg&Sari2016}, but all of them deposit most of the heat in the outer layers of the planet, well outside the metallic dynamo region.

As we will show in the Sect.~\ref{Sec: Argument for convection and dynamo suppression} and below, heat deposition above the dynamo region leads to significant differences. Further exploration of the heating radial profile has been conducted in detail in several other works, e.g. different power laws \citep{Ginzburg&Sari2015}, or Gaussian profiles at different depths \citep{Komacek&Youdin2017, Komaceketal2020}, and is not a matter of this study, where we only consider the two simplified scenarios above.

\begin{figure}[t]
\centering
\includegraphics[width=\hsize]{ 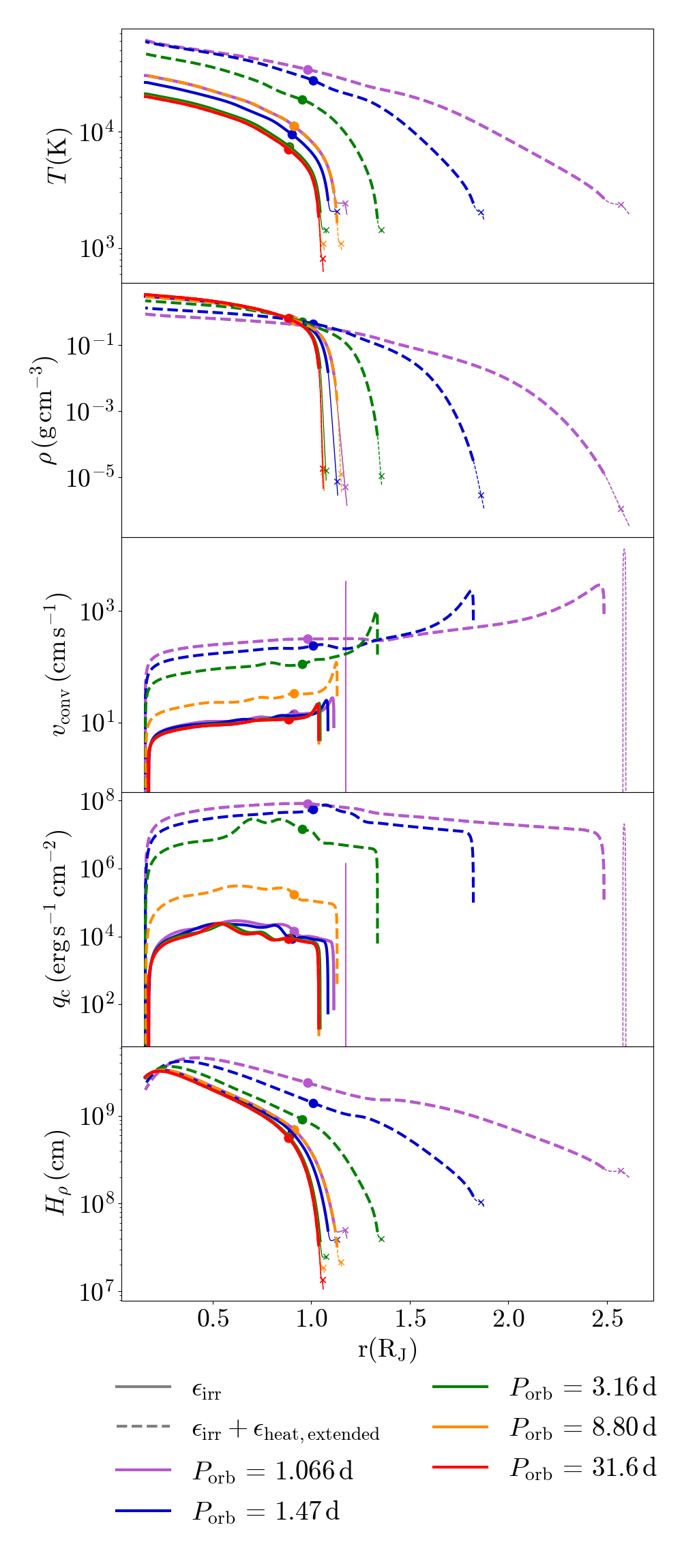}
\caption{Radial profiles of temperature, density, convective velocity, heat convective flux, and density scale height, for 1 $\mathrm{M_J}$ planets orbiting a 1 $\mathrm{M_\odot}$ star at 5 Gyr with different $P_{\rm orb}$. Models without and with extra heating (below $R_{\rm dyn}$) are shown by solid and dashed lines, respectively. $R_{\rm irr}$ is marked with a cross, $R_{\rm RCB}$ is the transition from thick to thin lines, and $R_{\rm dyn}$ is marked with a dot.}
\label{Fig: General HJ profiles}
\end{figure}

\subsection{Extended heat cases: general properties}
\label{Sec: General properties}

In Fig.~\ref{Fig: Radius as a function of orbit} we show the evolution of the planetary radius of a 1 $\mathrm{M_J}$ planet orbiting a 1 $\mathrm{M_\odot}$ star at 5 Gyr, as a function of the orbital period. In the plot, we compare irradiated models and heated models with the extended type of injection. Planets heated as shown in Eq.~(\ref{Eq: Thorngen and Fortney heat}) \citep{Thorngren&Fortney2018} inflate considerably and reach the range of observed HJ radii. The uncertainty for the outer radius of internally heated models reflects different values for $\epsilon_{\rm heat}(F_{\rm irr})$, considering the uncertainty ranges of the parameters in Eq.~(\ref{Eq: Thorngen and Fortney heat}). We also show a not-to-scale sketch of the different radii for illustration purposes. The order $R_{\rm P}>R_{\rm irr}>R_{\rm RCB}>R_{\rm dyn}$ always holds for both heated and non-heated models. There is only an exception for the highest irradiated cases, in which an additional convective outer layer above $R_{\rm irr}$ appears. More considerations regarding the behavior of convective regions for highly irradiated and heated models are found below.

In Fig.~\ref{Fig: General HJ profiles} we show the internal profiles for some chosen periods in Fig.~\ref{Fig: Radius as a function of orbit}. Specifically, we show the temperature $T$, the density $\rho$ the density scale height $H_\rho(r) = P /\rho g$, the convective velocity $v_{\rm conv}$, and the convective flux $q_{\rm c}$. The convective velocity comes from mixing length theory (MLT) based on the model in \cite{Kuhfuss1986}, which reduces to the expression given by \cite[chap. 14]{Cox&Giuli1968}, in the limit of long time steps (see \cite{Paxtonetal2011, Jermynetal2023} for details). We define $q_{\rm c}$ as the convective heat flux in the dynamo-generating region as follows:
\begin{equation}
    q_{\rm c} = \frac{2 \, c_P \, T \,~\rho^2 \, v_{\rm conv}^3}{P \, \delta}~,
    \label{Eq: Convective heat flux}
\end{equation}
where $\delta = - (\partial \, \mathrm{ln} \rho/ \partial \, \mathrm{ln} \, T)_P$.

When the planets are far from the heating efficiency peak, there is almost no difference between heated and non-heated models. Instead, inflation implies very different radial profiles: $T$, $H_{\rho}$, $v_{\rm conv}$, and $q_{\rm c}$ substantially increase for a given model and age, while the density in the innermost parts decreases due to the higher internal temperature (i.e., inflation and smaller mass/volume ratio). 

Note that we analyze models at 5 Gyr, which are representative of the bulk of HJs. Most evolution models of irradiated, internally heated HJs show that at $\sim$Gyr ages (corresponding to the vast majority of known HJs), the shrinking and cooling stalls, due to the essential balance between the internal heating and the long-term cooling (e.g., \citealt{Komacek&Youdin2017}). Therefore, the internal structure doesn't change notably. In reality, this behaviour relies on the assumption that both the stellar luminosity and the internal heat are constant in time, then neglecting the evolution of stellar luminosity (see \citealt{Lopez&Fortney2016, Komaceketal2020} for studies of re-inflation in this context), and the possible change in the heating efficiency of the underlying mechanism (see \cite{Viganoetal2025} for Ohmic models with evolving heating rates).

\subsection{Heat deposited in the outer region: suppression of convection}
\label{Sec: Argument for convection and dynamo suppression}

\begin{figure}[t]
\centering
\includegraphics[width=0.5\textwidth]{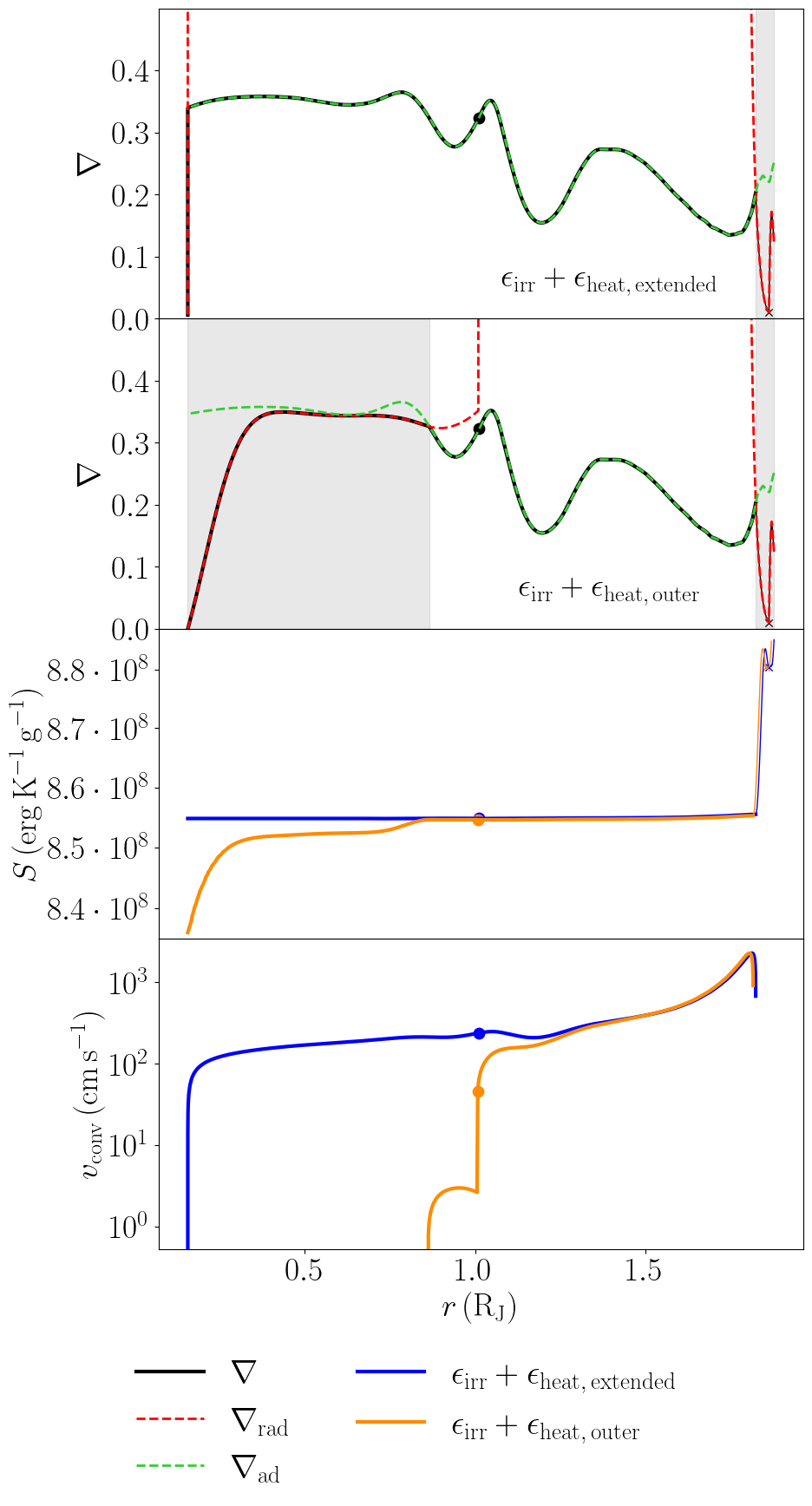}
\caption{Radial profiles of extended and outer heating models for a 1 $\mathrm{M_J}$ planet orbiting a 1 $\mathrm{M_\odot}$ star at 5 Gyr, with  $P_{\rm orb}$= 1.47 d. {\em Top panels}: Total logarithmic temperature gradient $\nabla$ (black solid line), with the adiabatic (green dashed line) and radiative (red dashed line) temperature gradients, $\nabla_{\rm ad}$ and $\nabla_{\rm rad}$, respectively. Gray zones mark stably stratified layers, i.e., $\nabla_{\rm rad}<\nabla_{\rm ad}$. {\em Bottom panels}: Specific entropy and convective velocity for the same models.}
\label{Fig: Extensive and outer heating comparison}
\end{figure}

Substantial differences appear with the outer type of heat injection. The heat is deposited in a significantly shallower layer, which, for a high enough deposited heat rate, leads to internal convection suppression. This is physically justified by comparing the logarithmic temperature radiative gradient ($\nabla_{\rm rad} = 3 \kappa L P/64\pi \sigma G m T^4$) and convective/adiabatic gradient ($\nabla_{\rm ad}$), which MESA determines using MLT. Following the Schwarzchild criterion, the actual temperature gradient  $\nabla$ is set to the smaller of the adiabatic or radiative, i.e., $\nabla= d \ln T/ d \ln P=\min(  \nabla_{\rm rad}, \nabla_{\rm ad})$. Therefore, convection develops where $\nabla_{\rm ad}$ is lower than $\nabla_{\rm rad}$, and on the other hand, the stably stratified layers arise when $\nabla_{\rm rad}$ is lower than $\nabla_{\rm ad}$. Similar to \cite{Komacek&Youdin2017}, we plot $\nabla, \nabla_{\rm rad}$ and $ \nabla_{\rm ad}$ in the two top panels of Fig.~\ref{Fig: Extensive and outer heating comparison} for the same 1 $\mathrm{M_J}$ planet with a $P_{\rm orb}$= 1.47 d. The gradient inversion is always present in the outermost part of the planets, covering at least the irradiation depth. For the model with outer heating, the inversion also happens in a substantial part of the interior near the core, creating an additional internal stratified layer.

This behavior can also be seen with the specific entropy gradient: the extended heated model (blue lines) remains isentropic over a large volume, whereas the outer heating model produces a positive entropy gradient in correspondence with the layers just below where the heat is deposited. With the outer heating case, $v_{\rm conv}$, and thus $q_{\rm c}$, greatly reduces and even vanishes below $R_{\rm dyn}$. The reduction of convective region size has been briefly mentioned in previous works where intense heating was deposited at specific interior locations \citep[e.g.,][]{Komacek&Youdin2017, Komaceketal2020}. Importantly, it has relevant consequences for the internal dynamo (see Sect.~\ref{Sec: Magnetic scaling laws}).

Note that, on the contrary, planetary radius, external luminosity (set by $R_{\rm P}$ and $F_{\rm irr}$), and the profiles of $T$, $\rho$, $P$, $g$ are almost indistinguishable between the two heating types. This is because the layers responsive to heating are the external ones, due to their lower density. They are the ones responsible for the inflation, as seen by comparing how the density profiles extend, in Fig.~\ref{Fig: General HJ profiles}. Therefore, depositing the heat over most of the planet, or only in the outer convective region, provides a very similar evolution of thermodynamic profiles. 

Additionally, in the most irradiated models, a shallow convective layer arises in the outermost regions of the planet. HJ shallow layer is conductive due to alkali thermal ionization \citep{Kumaretal2021} and thus could be potentially an additional source of convective dynamo. However, note that the convection is suppressed if the internally generated magnetic field is strong enough, higher than the critical value, $B_{\text{crit, r}}$, given by \cite[Eq. 9]{Jermyn&Cantiello2020}. Among the models shown in Fig.~\ref{Fig: General HJ profiles}, only the extremely irradiated case with $P_{\rm orb}\simeq 1$ d shows this shallow convective outer layer, as seen in both $v_{\rm conv}$ and $q_{\rm c}$ panels. Its associated $B_{\text{crit, r}}$ corresponds to about 100 G, which, as we will see in Sect.~\ref{Sec: Magnetic scaling laws}, is higher than the values that we estimate, therefore leaving open the possibility of such an additional convective layer. However, note that these results are sensitive to external boundary conditions and the heat deposition $\epsilon_{\rm irr}(P)$. Furthermore, the surface deposition for HJs is far from uniform, which would lead to a stronger convection layer in the dayside and hamper its existence in the nightside. Global circulation three-dimensional models are needed to address the possible existence of such a layer and their interaction with the thermal winds.

\begin{figure*}[ht!]
\centering
\includegraphics[width=\textwidth]{ 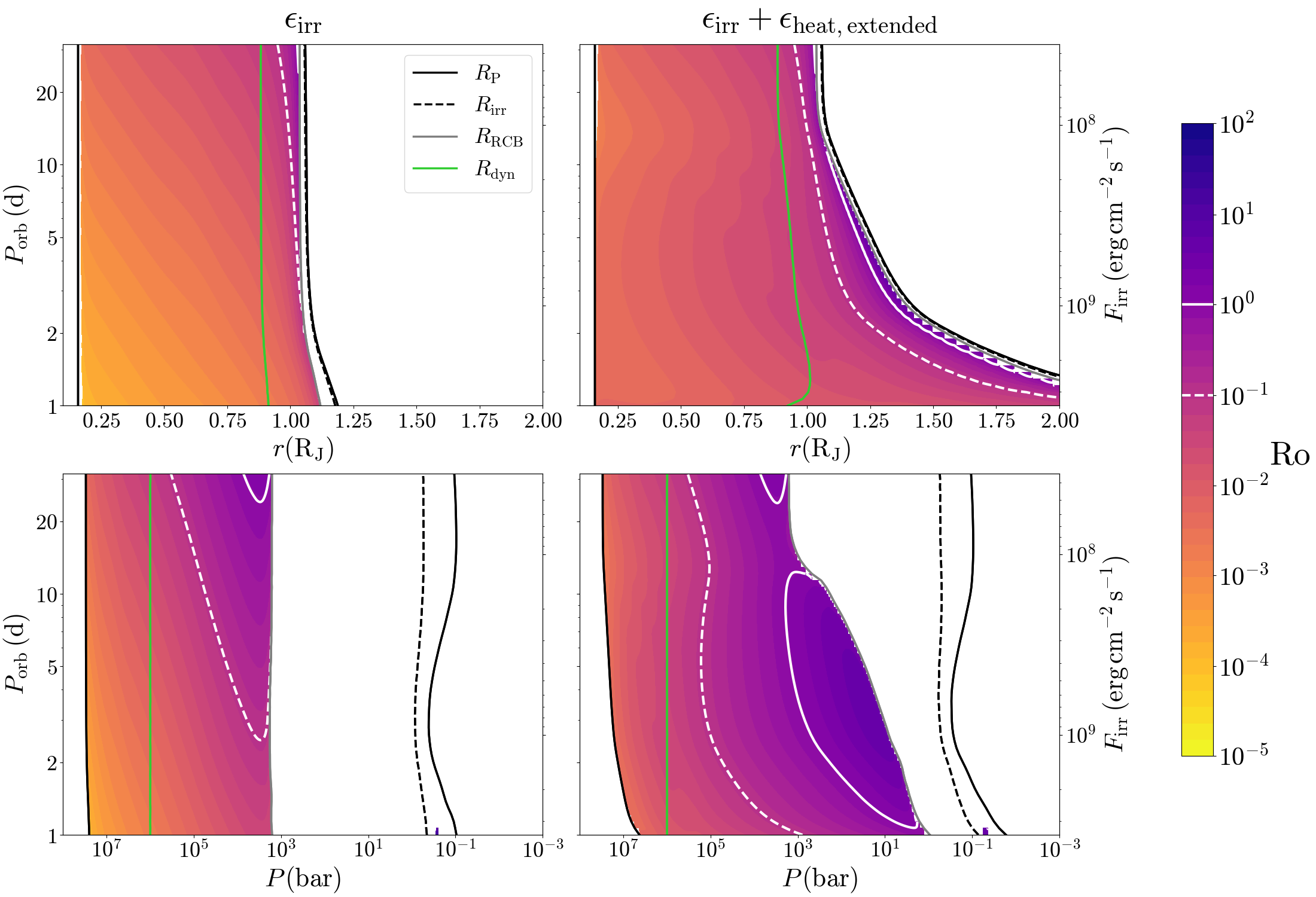}
\caption{Rossby number as a function of radius or pressure for 1 $\mathrm{M_J}$ planets orbiting a solar-like G2 star, comparing heated (left panels) and non-heated (right panels) models. The different radii introduced in Fig.~\ref{Fig: Radius as a function of orbit} are shown as in the legend, while the white lines mark $\mathrm{Ro}$=0.1 (dashed) and $\mathrm{Ro}$=1 (solid).}
\label{Fig: Rossby number as a function of distance}
\end{figure*}

\begin{figure}[ht]
\centering
\includegraphics[width=0.45\textwidth]{ 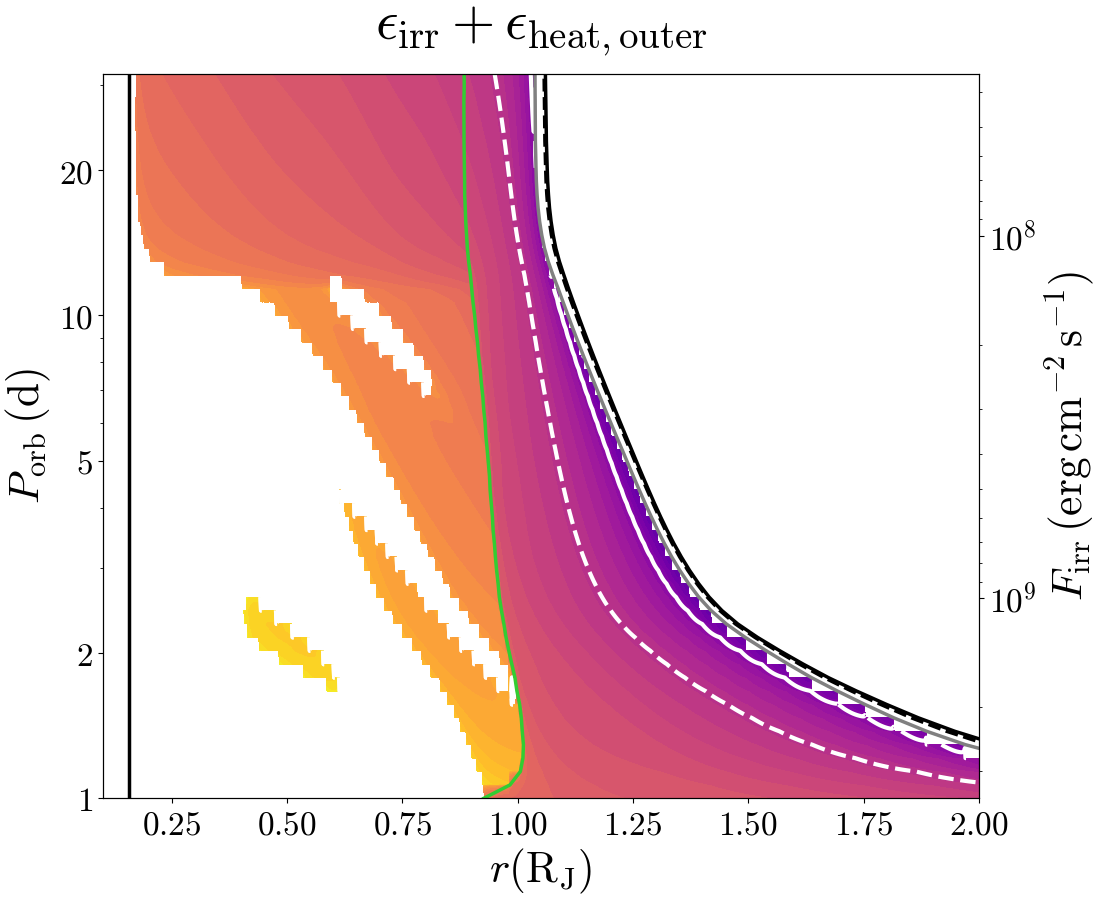}
\caption{Similar to Fig.~\ref{Fig: Rossby number as a function of distance}, but for the outer deposition of heat, i.e., only above the dynamo region.}
\label{Fig: Rossby number for external heating}
\end{figure}

\section{Rossby number estimation}
\label{Sec: Rossby number estimation}

The Rossby number, $\mathrm{Ro}$, is the ratio between inertial forces (dominated by convection) and the Coriolis force. Most planets in the solar system are fast rotators, i.e. they are in a low $\mathrm{Ro}$ regime: defining $\mathrm{Ro} = P_{\rm rot}/\tau_{\rm turn}$, i.e., the ratio of rotation period to convection turnover timescales, Earth and Jupiter have Ro $\sim 10^{-5}$ and $10^{-4}$, respectively. In this regime, the widely used dynamo scaling laws by \cite{Christensenetal2009}, explained below, apply.
In HJs, as seen in Sect.~\ref{Sec: Irradiation and tidal synchronization}, tidal locking implies $P_{\rm rot} \, = \, P_{\rm orb}$. We obtain the convective timescale, $\tau_{\rm turn}(r)$, from the evolutionary models, by using the radial profiles of density scale height, $H_\rho(r)$, and convective velocity, $v_{\rm conv}(r)$, so that
\begin{equation}
\mathrm{Ro}(r) = \frac{P_{\rm orb} \, v_{\rm conv}(r)}{H_\rho(r)}~.
\end{equation}
In Fig.~\ref{Fig: Rossby number as a function of distance} we show $\mathrm{Ro}(r)$, for 50 different values of $P_{\rm orb} \in [1,25]$ d, considering the representative case of a 1 $\mathrm{M_J}$ orbiting a 1 M$_\odot$ star, at 5 Gyr. We compare purely irradiated cases with the irradiated plus extended heated ones. Even though the extra heat mostly inflates the outer layers, it leads to $\mathrm{Ro}$ increasing by about 1 order of magnitude in the inner layers, due to the higher internal entropy. Still, the dynamo region maintains $\mathrm{Ro} < 0.1$, regardless of the stellar irradiation, so that we can infer that HJs remain in the fast-rotation regime. Since the evolutionary models depend on the irradiation alone, the results for Ro are the same for other stellar types, but shifted in orbital periods, as shown in App.~\ref{App: Stellar type dependence} for MV3, KV3, and FV5 stars. HJs around the most massive star lead to larger inflation and higher values for $\mathrm{Ro}$, but the dynamo region still remains lower than 0.1 in all cases. The sensitivity of results with planetary masses is discussed below.

In Fig.~\ref{Fig: Rossby number for external heating} we show the same upper right panel of Fig.~\ref{Fig: Rossby number as a function of distance} for the case with heat injection above the dynamo region. As expected from Fig.~\ref{Fig: Extensive and outer heating comparison}, there are large non-convective areas, where $\mathrm{Ro}$ is formally zero, in the metallic hydrogen region.

\begin{figure}[t]
\centering
\includegraphics[width=\hsize]{ 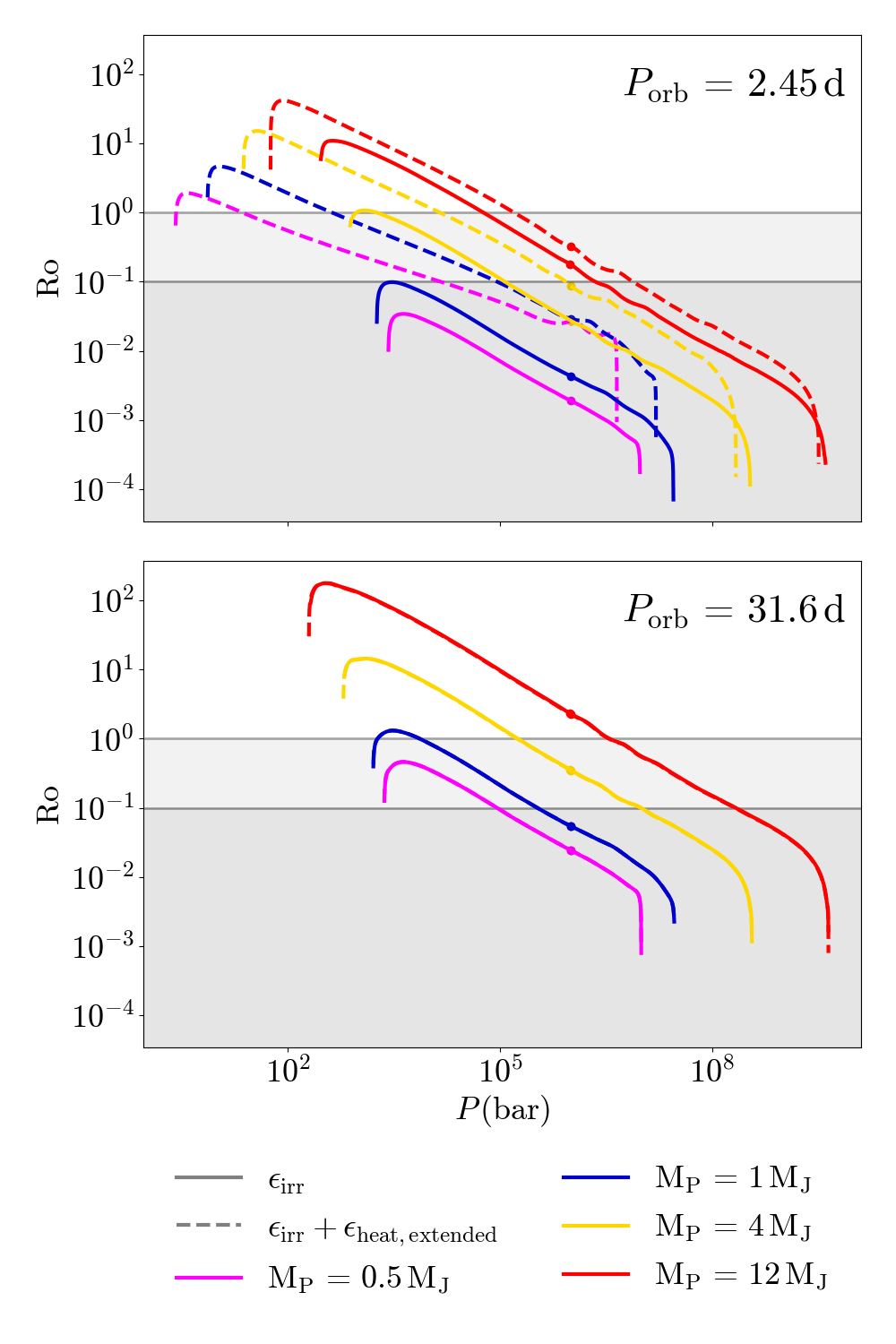}
\caption{Rossby number as a function of pressure for different mass planets orbiting a G2 star with two different periods, comparing models with the extended heating (dashed) with only irradiated models (solid).}
\label{Fig: Rossby number different mass planets}
\end{figure}

In Fig.~\ref{Fig: Rossby number different mass planets} we show $\mathrm{Ro}$ for several different mass planets, for two representative orbital periods $P_{\rm orb}=2.45$ and 31.6 d, for both heated and non-heated models. We show four planetary masses, which well represent the monotonic trend with $M_{\rm P}$. Heated and non-heated models are indistinguishable for long periods, as the $\epsilon(F_{\rm irr})$ is negligible. As seen in the top panel of Fig.~\ref{Fig: Rossby number different mass planets}, inflation has minimal effect for planets above 8~$\mathrm{M_J}$, with only a slight increase of $\mathrm{Ro}$ with a factor less than 2. Instead, a low-mass planet of 0.5~$\mathrm{M_J}$ experiences a more than one order of magnitude increase in $\mathrm{Ro}$, although it remains in the fast rotator regime. This still has consequences for the interior dynamics, as an increase in $\mathrm{Ro}$ may lead to more multipolar but weak dynamo-generated magnetic fields \citep{Christensen&Aubert2006, Jones2011, Davidson2013}. 

Planets with longer $P_{\rm orb}$ but still tidally locked, as the ones shown on the top panel of Fig.~\ref{Fig: Rossby number different mass planets}, have the highest $\mathrm{Ro}$. The most conductive layers of low-mass planets are generally well within the fast-rotation regime, but for masses higher than 4~$\mathrm{M_J}$, a large fraction of their metallic hydrogen regions have $\mathrm{Ro}\gtrsim 0.1$, being the only exceptions to the overall finding of this study. Note that for longer period planets, $P_{\rm orb}>15$ d, the heating mechanism is low and there is no inflation (see App.~\ref{App: Stellar type dependence}). Therefore, the argument for convection suppression would not apply to high-mass long-period planets. 

\section{Observational consequences}
\label{Sec: Observational consequences}

\subsection{Magnetic field scaling laws}
\label{Sec: Magnetic scaling laws}

\begin{figure*}[ht!]
\centering
\includegraphics[width=\hsize]{ 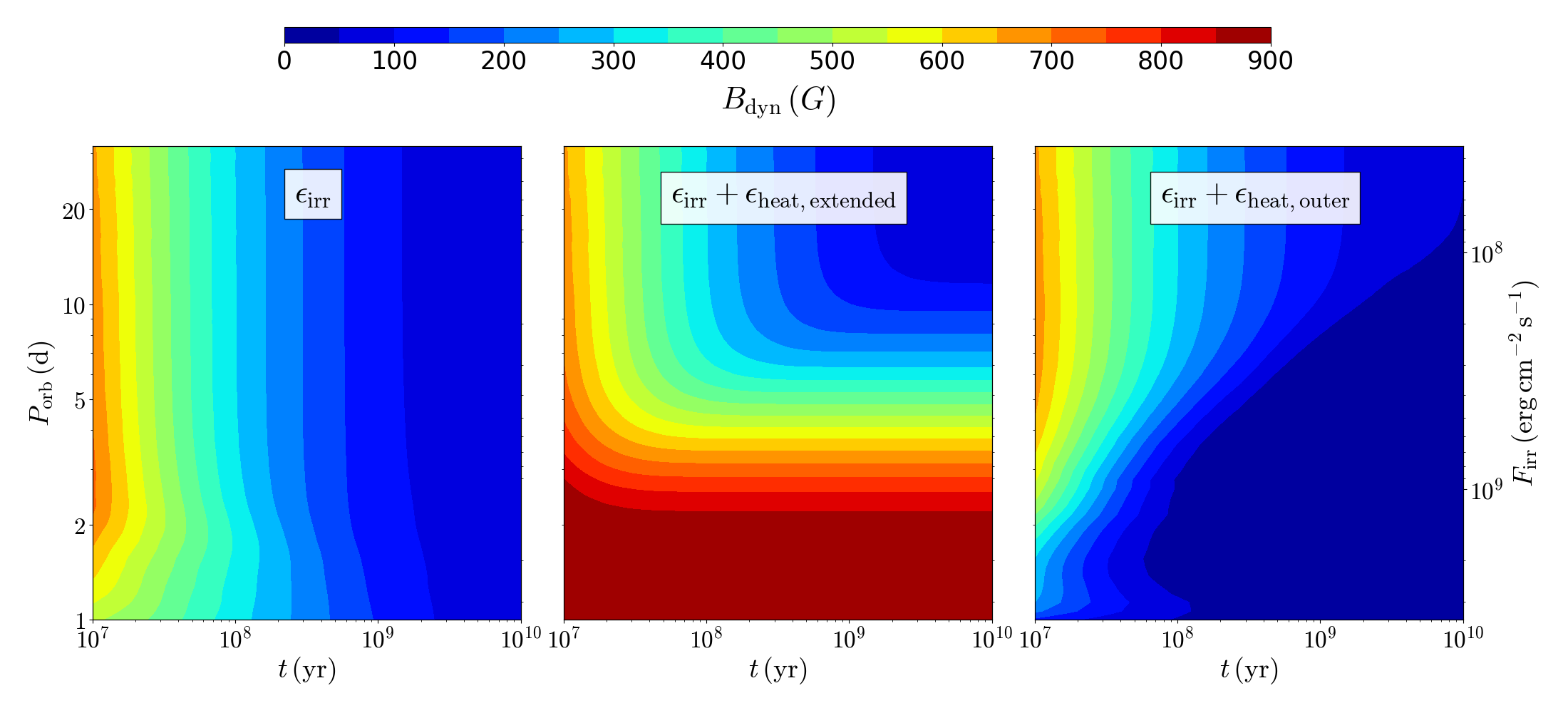}
\includegraphics[width=\hsize]{ 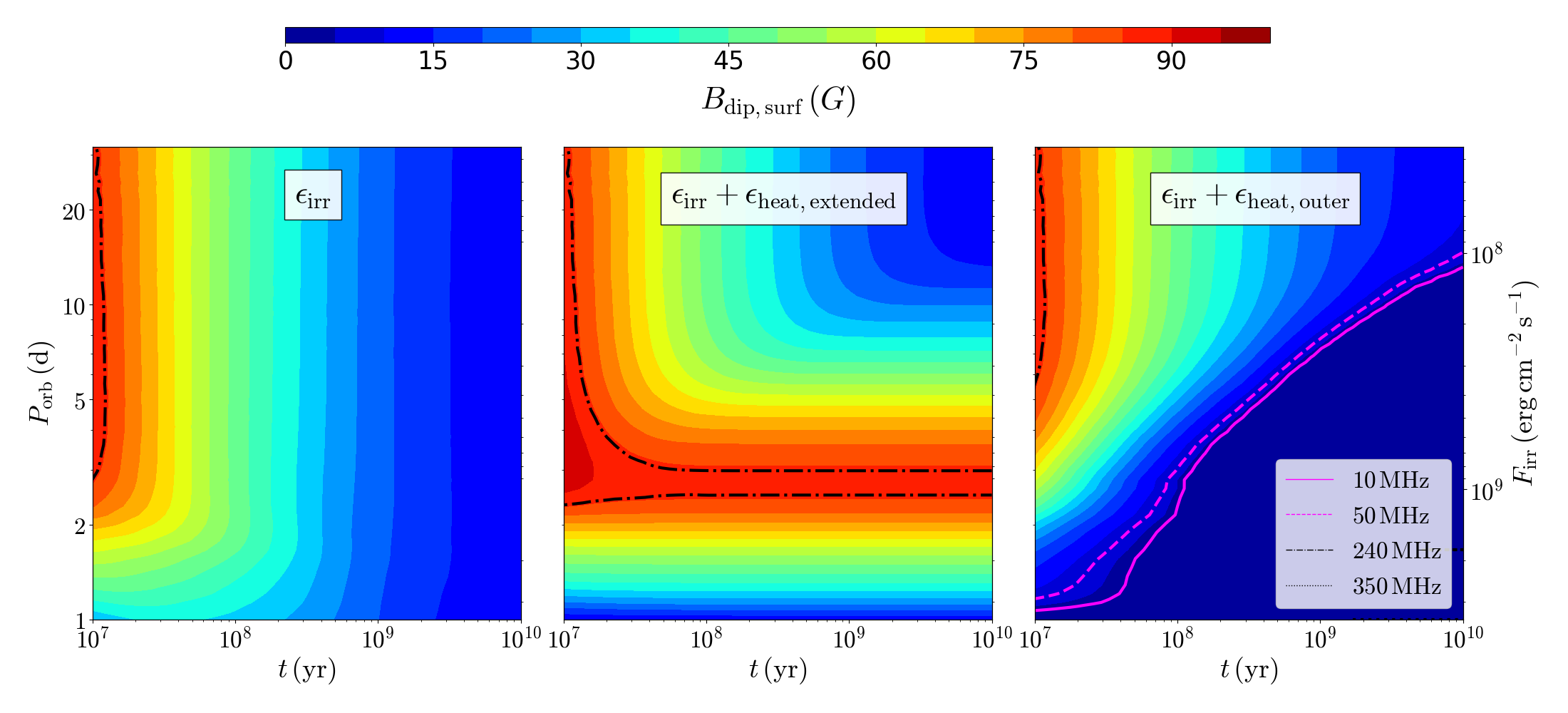}
\caption{Magnetic field estimates for 1 $\mathrm{M_J}$ planets orbiting a 1 $M_\odot$ main-sequence star (same models as Fig.~\ref{Fig: Rossby number as a function of distance} and~\ref{Fig: Rossby number for external heating}). {\em Top panels}: average magnetic field strength at the dynamo surface obtained with the scaling law discussed in the text. {\em Bottom panels}: inferred dipole component at the planetary surface equator. Several of the most interesting frequency limits are included: 10 MHz, the ionospheric cut-off and lower range of LOFAR; 50 MHz, the lower range of SKA-low; 240 MHz, the upper range of LOFAR; and 350 MHz, the upper range of SKA-low.}
\label{Fig: HJ magnetic field scaling laws}
\end{figure*}
Combining energy equipartition arguments, the results for a large set of dynamo simulations \citep{Christensen&Aubert2006} and magnetic field measurements, \cite{Christensenetal2009} argued that for the Earth, Jupiter, and fully-convective fast-rotating main-sequence stars, the convective heat flux is the key factor determining the order of magnitude of the dynamo-generated magnetic field strength:
\begin{equation}
    \frac{B^2}{2 \mu_0} \propto f_{\rm ohm} \,~\rho^{1/3} \, \left(\frac{q_{\rm c}L}{H_T}\right)^{2/3}~,
    \label{Eq: Bdyn scaling law general}
\end{equation}
where $\mu_0$ is the magnetic permeability, $f_{\rm ohm}\leq1$ is the ratio of ohmic to total dissipation, $L$ is the length scale of the largest convective structures and $H_T =  P / (\rho g \nabla_{\rm ad})$ is the temperature scale height, with $c_p$ the heat capacity, $\alpha$ the thermal expansion coefficient, $g$ is the gravitational acceleration, and $q_{\rm c}$ is the convective heat flux as previously defined. As in \cite{Christensenetal2009}, we hereafter assume $f_{\rm ohm}=1$ for simplicity. Eq.~(\ref{Eq: Bdyn scaling law general}) can be integrated in the spherically symmetric shell of volume $V$ (between $R_{\rm core}$ and $R_{\rm dyn}$) and define the root-mean-square value of the magnetic field in the dynamo region as:
\begin{equation}
    \frac{B_{\rm dyn}^2}{2 \mu_0} := \frac{\langle B~\rangle^2}{2 \mu_0} = c \, f_{\rm ohm} \, \langle~\rho~\rangle^{1/3} \, \left( F \, q_o~\right)^{2/3},
    \label{Eq: Bdyn scaling law real}
\end{equation}
where brackets indicate volume averages, $q_0$ is a reference convective flux, which we take as the value of $q_{\rm c}$ at $R_{\rm dyn}$, $c$ is a proportionality constant, calibrated as $c=0.63$ by \cite{Christensenetal2009}, and the factor $F$ includes the radial profile variations as follows:
\begin{equation}
    F^{2/3} = \frac{1}{V} \int_{R_{\rm core}}^{R_{\rm dyn}} \left( \frac{q_{\rm c}(r)}{q_0} \frac{L(r)}{H_T(r)}~\right)^{2/3} \left( \frac{\rho(r)}{\langle~\rho~\rangle}~\right)^{1/3} 4 \pi r^2 \, dr
    \label{Eq: F scaling law}
\end{equation}
where $L$ = min$(D, H_\rho)$. All these quantities are obtained for the evolutionary models at each timestep. This implementation is similar to other planetary studies \citep[e.g.][]{Hori2021, Kilmetisetal2024}. 

Additionally, we can estimate the dipolar component at the planetary surface at the magnetic equator, $B_{\rm dip, surf}$, defined as: 
\begin{equation}
    B_{\rm dip, surf} = \frac{1}{2\sqrt{2}} \left( \frac{R_{\rm dyn}}{R_{\rm P}}~\right)^3 B_{\rm dyn}~,
    \label{Eq: Bdip}
\end{equation}
where for simplicity we use the same factor $1/(2\sqrt{2})$ as in \cite{Reiners&Christensen2010}, which comes from the assumption that the rms dipole field strength is half of the rms and that we consider the value at the magnetic equator, which is 1/$\sqrt{2}$ of the average dipole field. Note also that the scaling law suffers of considerable uncertainties: (i) \cite{Reinersetal2009, Reiners&Christensen2010} reformulate the original scaling law and evaluate a $\sim 60\%$ relative dispersion around that formula from data comparison; (ii) several factors (like the dipolar fraction and $f_{\rm ohm}$) which may be in reality systematically shifted, or vary from case to case; (iii) the calibration of the scaling law has been originally limited to the Earth, Jupiter and low-mass main-sequence stars, excluding other planets, with brown dwarfs found later to be surprising outliers \citep{Kaoetal2016, Kaoetal2018}, and with no available comparison with magnetic field measurements for exoplanets, except a few indirect estimates by \citep{Cauleyetal2019}. Therefore, the absolute values should be taken with care. Nevertheless, our main focus is to show the trends of the predicted field, mainly as a function of age, $P_{\rm orb}$, and the type of heating.

Note also that, for each model, we show the evolution from 10 Myr to 10 Gyr. At early ages, convection is more vigorous, but more effects related to disk-planet interaction and migration could occur, which we do not account for. Moreover, as discussed above, the internal structure predicted by cooling models still depends on the specific initial entropy, while at later ages, $\gtrsim 0.1$ Gyr, the evolution has lost memory of the initial condition, so that results are more reliable in this sense. Since the bulk of detected HJs are ${\cal O}$(Gyr)-old, hereafter we will not focus on the $t\lesssim 100$ Myr part of the plots.

In Fig.~\ref{Fig: HJ magnetic field scaling laws}, we show $B_{\rm dyn}$ (top) and $B_{\rm dip, surf}$ (bottom) for the same 1 $\mathrm{M_J}$ models shown in Fig.~\ref{Fig: Rossby number as a function of distance} and~\ref{Fig: Rossby number for external heating}. Models with only irradiation (left panels) lead to a magnetic field decay in time ($B \propto t^{-0.3}$, where $B$ can be $B_{\rm dip, surf}$ or $B_{\rm dyn}$) proportional to the gradual cooling of the planet and independent of the orbital period, compatible with \cite{Reiners&Christensen2010} and \cite{EliasLopezetal2025}. Models with internal heat lead instead to much higher $B_{\rm dip, surf}$, since the shrinking is stalled, with an equilibrium between the long-term cooling and the internal heating. These values are compatible with \cite{Yadav&Thorngren2017}, which assumes that all the heat necessary to inflate the planet is involved with the magnetic field generation (see \citealt{Viganoetal2025} for a discussion about the assumptions in the use of the scaling laws). As seen in Fig.~\ref{Fig: Radius as a function of orbit} and Fig.~\ref{Fig: General HJ profiles}, inflation mostly affects the outer layers of the planet, and the dynamo region does not expand nearly as much. The consequence is that, even though $B_{\rm dyn}$ plateaus at very high values $\sim 900$~G for $P_{\rm orb}\lesssim2$ d, $B_{\rm dip, surf}$ has much lower values, since the layers between the outer dynamo region $R_{\rm dyn}$ and the planetary surface $R_{\rm P}$ are inflated, and the relative decay of the dipolar field, cubic in radius, is relevant (higher multipoles decay even more). The non-heated models shown in \cite{Kilmetisetal2024} have a similar trend where the magnetic field estimation decays for short $P_{\rm orb}$.

The predictions for models with the outer heating, for which convection can be suppressed, are very different, as shown in the right panels of Fig.~\ref{Fig: HJ magnetic field scaling laws}. Whenever inflation becomes noticeable, the magnetic field strength dramatically decreases to values even lower than the Jovian values of $\sim$ 10 G. The cases with the lowest predicted values of $B_{\rm dyn}$ have thin layers of convection below $R_{\rm dyn}$. This implies that the integral, Eq. (\ref{Eq: F scaling law}), is very sensitive to both the assumed definition of $R_{\rm dyn}$, and the radial interval used in the outer heating. Moreover, it might not be accurate to estimate the magnetic field strength created in these shallower convective layers with the same scaling laws. Therefore, the values shown on the left panel of Fig.~\ref{Fig: HJ magnetic field scaling laws} could be even lower or potentially vanish for the convection-suppressed cases.

This fact leads us to draw a parallel with Venus. The most accepted argument for the lack of an active Venusian dynamo is the presence of a stagnant lid. The absence of an active tectonic plate system hinders the efficient cooling of its core, resulting in steeper entropy gradients \citep{Nimmo2002, Jacobsonetal2017}. The argument works similarly for HJs: if the heating necessary to reach the observed HJ radii is deposited in the outer layers, convection (and thus dynamo) is suppressed, essentially due to a blanketing effect.

\begin{figure*}[ht!]
\centering
\includegraphics[width=.95\hsize]{ 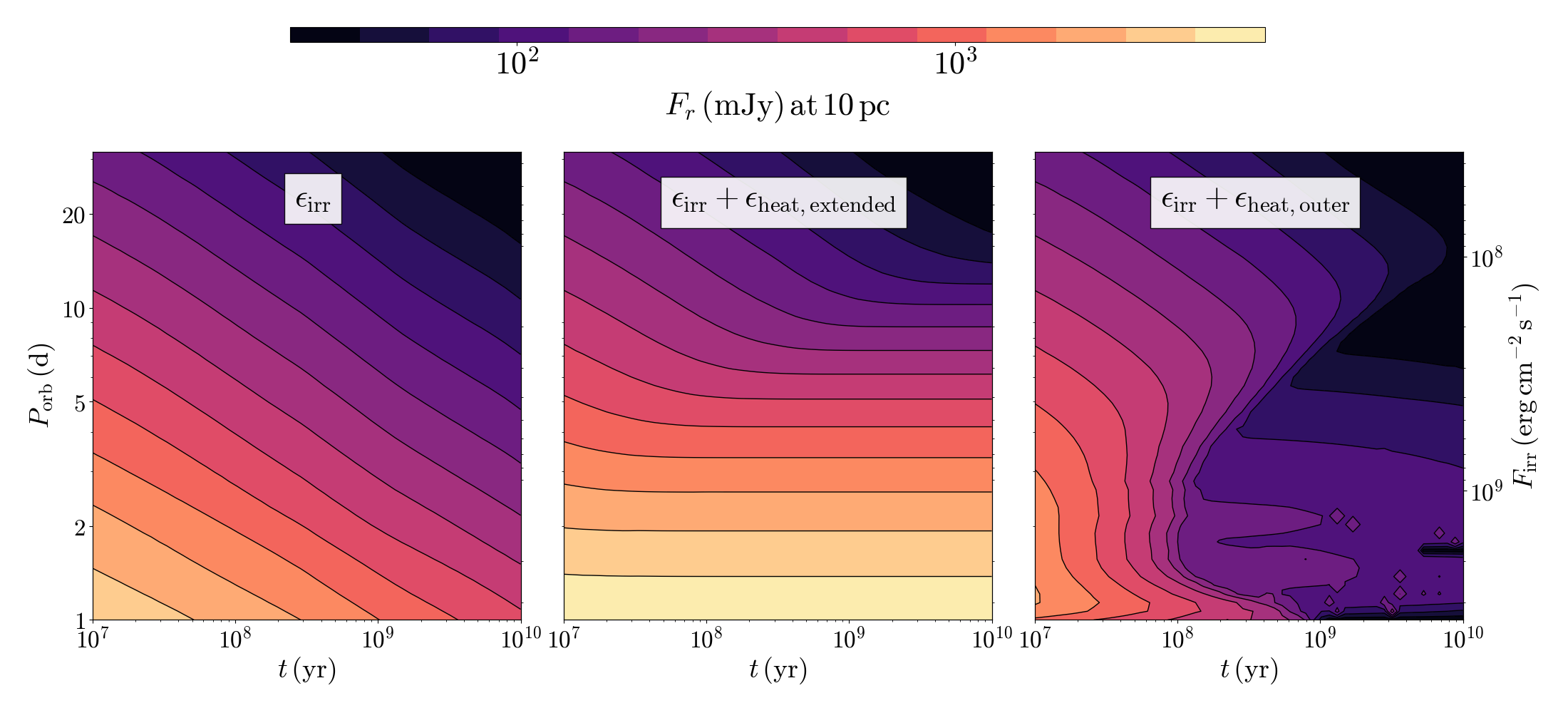}
\includegraphics[width=.95\hsize]{ 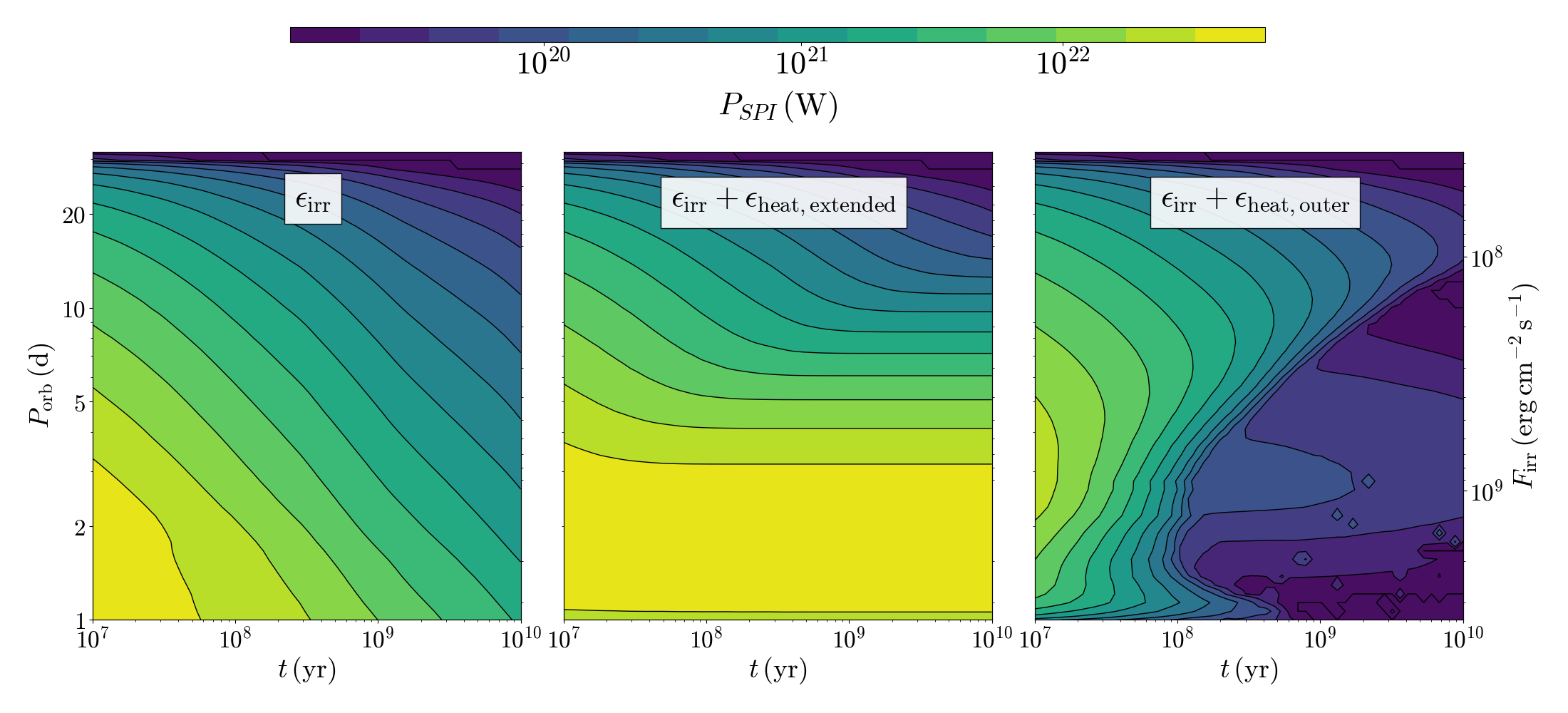}
\caption{Estimates for Jovian-like coherent radio flux (top, for a 10 pc away HJ) and SPI available power (bottom), as a function of age and $P_{\rm orb}$, for the same models shown in Fig.~\ref{Fig: HJ magnetic field scaling laws}.}
\label{Fig: HJ radio emission}
\end{figure*}

\subsection{Planetary coherent radio emission}
\label{Sec: Coherent radio emission}

One of the main observational consequences of exoplanetary magnetic fields is the possible coherent radio emission, as seen in all magnetized planets in the Solar system \citep{Zarka1998}. The mechanism behind this emission is the above-mentioned ECM \citep{Dulk1985}, which requires the presence of a source of keV electrons (arguably coming from the stellar wind) and a planetary magnetic field. As the charged particles from the stellar wind come into the vicinity of the planetary magnetic field, they start following helical trajectories around the magnetic field lines and get bunched together in phase. This results in a coherent, circularly polarized radio emission from the planet in a hollow cone, at the Larmor frequency of the local magnetic field $B$, $\nu_c$ [MHz] $ = 2.8 B$ [G]. Therefore, the detected emission traces the intensity of the magnetic field, with a frequency cut-off at the Larmor frequency of the largest value of the field in the emitting region, i.e., close to the surface. As seen in the bottom panel of Fig.~\ref{Fig: HJ magnetic field scaling laws}, irradiated and extended heated models have $B_{\rm dip, surf}$ estimations that surpass the $\sim$ 10 MHz ionospheric cut-off, which corresponds to a minimum magnetic field strength of $\sim$ 3.5 G. On the contrary, when HJ are inflated with the outer heating, their $B_{\rm dip, surf}$ usually are below the ECM cut-off, suppressing any detectable radio emission.

Given that the Jovian case remains the only firm observable constraint for gas giants' ECM emission, there are huge uncertainties on how bright the expected radio emission should be. Simple estimates assume that the emitted radio power is proportional to the kinetic flux intercepted by the planetary magnetosphere and is given by the phenomenological radiometric Bode’s law, which fits the well-characterized radio luminosity of Jupiter and the one from the other solar planets, which are less studied, coming mostly from limited in-situ Cassini and Voyager measurements. The corresponding specific flux, scaled by the measured typical Jovian values, reads \citep{Bastianetal2000, Zarkaetal2001, Stevens2005}:
\begin{equation}
\begin{aligned}
    F_r = 2.35 \cdot 10^{-2} \, \mathrm{mJy} \, \cdot \left( \frac{\dot{M}_\star}{\dot{M}_\odot}~\right)^{2/3} \cdot \left( \frac{M_{dip}}{M_{dip,J}}\right)^{2/3} \cdot \\ \cdot \left(\frac{a}{5 \, \mathrm{AU}}\right)^{-4/3} \cdot  \left(\frac{V_W}{400 \, \mathrm{km \, s^{-1}} }\right)^{5/3} \cdot \left(\frac{d}{10 \, \mathrm{pc}}\right)^{-2}
    \label{Eq: Radio emission}
\end{aligned}
\end{equation}
where $\dot{M}_\star$ is the stellar mass loss rate, $M_{dip} = B_{\rm dip, surf} \, R_{\rm P}^3$ is the planetary magnetic dipolar moment, and $d$ is the distance to the source. The stellar wind properties may vary from case to case and can be inferred from UV and X-ray observations (see e.g. \citep{Woodetal2021} for M dwarfs), or some models (e.g., \citealt{Johnstoneetal2015}). Given the intrinsic uncertainties, here we simply assume a representative solar value, $\dot{M}_\star=\dot{M}_\odot \simeq 400$ km s$^{-1}$ \citep{Woodetal2005}, thus neglecting the large spatial and time variability measured for the Sun (e.g., Fig. 1 of \citealt{Johnstoneetal2015}), and its dependence on $a$ from measurements or from e.g. the \cite{Parker1965} model.

In the top panel of Fig.~\ref{Fig: HJ radio emission}, we show the resulting estimates for the ECM radio flux, for the same models in Fig.~\ref{Fig: HJ magnetic field scaling laws}. HJs with extended heating have very high flux predictions, overcoming 1 Jy, which is incompatible with much lower upper limits reported in literature for tens of HJs (e.g. \citealt{Narangetal2024} and references within). Models with external heating have instead flux estimates which are orders of magnitude smaller, especially for the hottest cases at Gyr ages, where the convection and dynamo suppression bring the most relevant effects. Moreover, they are mostly below the 10 MHz ionospheric cut-off shown in Fig.~\ref{Fig: HJ magnetic field scaling laws} and cannot be detected by ground-based radio interferometers.

Note that the absolute values of the estimated flux for the purely irradiated case are compatible with the cases proposed in the original works by e.g. \cite{Stevens2005}, and fluxes $F_r\gtrsim 0.1$ mJy should be easily detectable by targeted observations and/or current wide-field radio surveys at low frequencies, LOFAR Stokes V survey V-LoTSS in particular \citep{Callinghametal2023}. However, after two decades of radio campaigns targeting tens of theoretically promising HJs have led to no confirmed planetary radio emission (see e.g. Fig. 6 of \citealt{Narangetal2024}, and references within), with a claim for one case \citep{Turneretal2021}, not confirmed by an extended follow-up \citep{Turneretal2024}. Therefore, the models are probably very optimistic, in part because they are often calibrated to the peak values of the Jovian short bursts, rather than the average one. In any case, the purpose of this work is to discuss the trends with irradiation and heating models, rather than the absolute values, which are arguably plagued by huge intrinsic uncertainties.

\subsection{Star-planet interaction available power}
\label{Sec: SPI power}

Another potentially observable signature of planetary magnetism is the effect of magnetic SPI. One of the possible SPI mechanisms for HJ systems can arise when a close-in giant planet moves through its host star’s magnetic field, enabling energy transfer due to the stretching of magnetic flux tubes that connect the planetary and stellar magnetospheres \citep{Lanza2013}. This interaction can produce observable signatures, such as spectroscopically detectable modulation with the orbital period of chromospheric activity (in particular, the emission line Ca II K, \citealt{Cauleyetal2019}), or coherent emission at radio frequencies close to the stellar surface (e.g., \citealt{Sauretal2013, Perezetal2021}). Here we don't enter into detail of the specific manifestation of the SPI at different wavelengths, but we simply follow \cite{Lanza2013} to derive the total power available:
\begin{equation}
    P_{\rm SPI} \simeq \frac{2\pi}{\mu_o} \, f_{\rm AP} \, R_{\rm P}^2 \, (2B_{dip, surf})^2 \, v_{\rm rel},
    \label{Eq: SPI Lanza2013}
\end{equation}
where $v_{\rm rel} = 2 \pi a /P_{\rm rel}$ is the relative velocity between the planet and the stellar magnetic field lines (therefore, $P_{\rm rel} = 2\pi/(\omega_{orb}-\omega_\star)$, where $\omega_i$ are orbital frequencies), $2B_{\rm dip, surf}$ is the polar magnetic field (twice its equatorial value defined in Eq.~\ref{Eq: Bdip}), and $f_{\rm AP}$ is the fraction of the planetary surface covered by stellar magnetic flux tubes:
\begin{equation*}
    f_{\rm AP} = 1 - \left(1- \frac{3 \xi^{1/3}}{2+\xi}~\right)
\end{equation*}
where $\xi = B_\star(a)/(2B_{\rm dip, surf})$. $B_\star(a)$ is the stellar magnetic field at the given orbital separation, which we derive assuming a solar-like value at the stellar surface for simplicity, $B_\star$ = 10 G. Note that this underestimates the stellar field for at least M dwarfs, which, being fully convective, usually show hundreds to thousands gauss fields \citep{Reinersetal2022}, and for the highly magnetic, chemically peculiar Ap and Bp stars. However, the bulk of HJs orbit Sun-like stars whose spectral type is usually associated with (at most) Sun-like large-scale magnetic field values (e.g., \citealt{Seachetal2020}), therefore justifying our simple assumption.

Eq.~(\ref{Eq: SPI Lanza2013}) estimates the Poynting flux generated by the continuous stretching of magnetic field lines due to the planet’s motion and was also used in \cite{Cauleyetal2019} to estimate the magnetic field of four HJs by observing periodic enhancements in the Ca II K chromospheric emission line. In the same work, they show that other SPI estimates, like the pioneering reconnection model \citep{Cuntzetal2000} and the Alfv\'en wing scenario \citep{Sauretal2013, Strugareketal2016}, have different dependencies with the planetary and stellar magnetic fields, and typically provide less available power, incompatible with the observed activity modulation signal \citep{Cauleyetal2019}, so we don't discuss them here. In the bottom panels of Fig.~\ref{Fig: HJ radio emission}, we show the resulting available SPI power, a fraction of which (e.g., $\lesssim 1\%$ in \citealt{Cauleyetal2019}) can become visible as orbit-modulation of the activity indicators. Similar to Fig.~\ref{Fig: HJ magnetic field scaling laws}, only the extended heat scenario, for the hottest cases, provide optimistic predictions for inflated HJs, $P_{\rm SPI}\gtrsim 10^{22}$ W, potentially able then to power the observationally inferred values of $10^{20}$ W in Ca II K line modulation, assuming reasonable $\lesssim 1\%$ conversion efficiency, \citealt{Cauleyetal2019}. When the heat is applied only above the dynamo, the available power is typically reduced by orders of magnitude. 

Note that an essential requirement for SPI is that the local wind velocity should be less than the Alfv\'en velocity (i.e., the Alfvén Mach number $M_{A} = v_{W}/v_{alf}<1$, where $v_{alf}$ is the Alfvén speed). This intrinsically depends on the individual stellar properties, and we cannot assess it here. However, it can further limit the detectability of SPI signals. Therefore, again, the values here indicated should be regarded as upper limits and be used to only infer the trends with the HJ properties (depending ultimately on irradiation and heating).

\section{Conclusions}
\label{Sec: Conclusions}

In this work, we model the interior evolutionary tracks of inflated hot Jupiters using the one-dimensional evolutionary code MESA. We assume tidal locking and irradiation from main-sequence host stars, and we explore dependencies on orbital distance, planetary and stellar mass, and the type of heat injection. Guided by observationally constrained flux-heating efficiency relations \citep{Thorngren&Fortney2018}, we inject heat into the internal layers of the planet to reproduce the observed radii. This relation and others in the literature (e.g. \cite{Sarkisetal2021}) have big dispersions around average values of inflated radii, for a given irradiation. Therefore, this study only addresses the trends with the planetary properties (mass, separation, type of internal heating), keeping the parameters involving the amount of deposited heat fixed.

For a given stellar type, orbital period and planetary mass, we compare the purely irradiated case with two simplified scenarios of regions where extra heat is continuously deposited: extended (injected mostly in the dynamo region, $P\gtrsim 10^6$ bar), or outside the dynamo region, which is what is expected in most models, the Ohmic one especially \citep{Batygin&Stevenson2010, Batyginetal2011, Thorngren2024, Viganoetal2025}. Whether heat is injected uniformly, centrally, or throughout the convective region, the structural differences are minimal. Planets typically exhibit a shallow stratified outer layer, which fully contains the irradiation zone, followed by a deep convective interior where pressure is sufficient for hydrogen metallization and possible dynamo action.

However, important differences are seen in the Rossby number as a function of depth. The vast majority of models yield $\mathrm{Ro}\lesssim 0.1$, indicating a fast-rotating convection regime. Defining HJs as having $P_{\rm orb}<$ 10 d or $T_{\rm eq}>$ 1000 K (see \cite{Ganetal2023} and within), we can safely say that the vast majority of confirmed HJs are expected to be in the low-$\mathrm{Ro}$ regime. Only the most massive, distantly orbiting (yet still tidally locked) planets exceed $\mathrm{Ro}\gtrsim 0.1$ over significant interior regions, potentially altering the dynamo regime. Thus, massive HJs with orbital periods beyond 15-20 days may host low-Rossby-number dynamos that generate weaker, more multipolar magnetic fields despite similar convective power. Within the known HJs, only three targets with $M_P\gtrsim 4~M_J$ closer than 100 pc have 15 d$<P_{\rm orb}<$40 d, all orbiting G-type stars. The most promising candidate to test their potential peculiarity is GJ 86 b \citep{Stassunetal2017}, which has a 15.8 d orbit, a mass of 4.4 $\mathrm{M_J}$, and is located only at 10.8 pc. The other two planets are farther away: HD 72892 b (39.4 d, 5.5 $\mathrm{M_J}$, 69.7 pc, \cite{Fengetal2022}) and TIC 393818343 b (16.2 d, 4.3 $\mathrm{M_J}$, 93.7 pc, \cite{Sgroetal2024}).

The extra heat affects the outer layers and consistently increases $\mathrm{Ro}$. While this effect is negligible for planets above 8~$\mathrm{M_J}$ (especially those already exceeding $\mathrm{Ro}\gtrsim 0.1$), which can hardly get inflated due to their higher gravity, lower-mass planets experience a roughly one order of magnitude increase in $\mathrm{Ro}$, which has important interior dynamic consequences even it they remain within the low-$\mathrm{Ro}$ regime. When the heat injection is localized outside the dynamo region, the temperature gradients are reduced or even inverted, as already briefly mentioned by e.g. \citep{Komacek&Youdin2017, Komaceketal2020}. The internal heat transport is significantly reduced, leading to positive entropy gradients that suppress convection. 

The relevant consequence of this comes when one applies the widely-used, observationally constrained magnetic scaling laws from \cite{Christensenetal2009}, suited for fast rotators. When heat is deposited in an extended way, we recover surface magnetic field strengths around 100~G for the most inflated planets, an order of magnitude higher than Jupiter. These estimates are compatible with the work of \cite{Yadav&Thorngren2017}, who assume that the convective heat flux is given by the extra heat. This is conceptually similar to our extended heat case, where heat is essentially deposited in the dynamo region. In contrast, a more realistic external heating substantially reduces convective power and yields weaker magnetic fields. 

This result is particularly interesting in the context of the Ohmic dissipation models \citep{Batygin&Stevenson2010, Batyginetal2011, Pernaetal2010b, Wu&Lithwick2013, Ginzburg&Sari2016, Knierimetal2022, Viganoetal2025}. Essentially, it relies on the presence of conducting outermost layers due to thermal ionization (e.g., \citealt{Kumaretal2021, Dietrichetal2022}), and on the circulation of strong, mainly zonal winds which twist and stretch the background field coming from the deep-seated dynamo. The inducted currents propagate and dissipate into the interior as well, due to the finite values of conductivity in the outer convective region (much smaller than in the dynamo region, but not zero). Combining estimation for the conductivity and the induced currents, models predict that the Ohmic heating profile steeply decays towards the interior, so that the bulk of it is deposited well above the dynamo region \citep{Batyginetal2011, Ginzburg&Sari2016, Knierimetal2022, Viganoetal2025}. Since the atmospheric induction depends also on the intensity of the background magnetic field (in a linear way only for magnetic Reynolds numbers Rm$\lesssim 1$), there could be an interesting, non-trivial coupling between the dynamo and the atmospheric induction: if the Ohmic heating is large, it can suppress convection and the underlying dynamo field, which would in turn decrease the Ohmic heating rate, causing the convection to be restored. This process might only lead to a less effective Ohmic heating mechanism or, potentially, show an oscillatory behavior, as proposed by \cite{Viganoetal2025}.

The considerations on the magnetic field strengths are of direct interest for the observational detection of SPI signals \citep{Cauleyetal2019}, and/or planetary Jovian-like coherent radio emission, which remains elusive, despite optimistic predictions \citep{Stevens2005} and tens of observational campaigns (e.g., \citealt{Narangetal2024} and references within). Several potential causes can account for the lack of detection, including the beaming effect, large distances, intrinsic variability, and limited observational coverage. However, one cannot neglect the possibility that the typical estimates of HJ magnetic fields of the order $100$ G might be too optimistic \citep{Yadav&Thorngren2017, Kilmetisetal2024}. In this sense, our results indicate that very massive HJs might offer an exception, since their internal structure is less sensitive to the combined effects of irradiation and heating, and could maintain large magnetic fields. For Jupiter-like masses, if the external heating model is assumed, there is little hope of having a magnetic field much larger than Jupiter.

\section*{Acknowledgements}
AEL, SK and CSG's work has been carried out within the framework of the doctoral program in Physics of the Universitat Autònoma de Barcelona. AEL, DV, SK and CSG are supported by the European Research Council (ERC) under the European Union’s Horizon 2020 research and innovation program (ERC Starting Grant "IMAGINE" No. 948582, PI: DV) and the support from the ``Mar\'ia de Maeztu'' award to the Institut de Ciències de l'Espai (CEX2020-001058-M). AEL gratefully acknowledges the support and hospitality from the Simons Foundation through the predoctoral program at the Center for Computational Astrophysics, Flatiron Institute. The computational resources and services used in this work were provided by facilities supported by the Scientiﬁc Computing Core at the Flatiron Institute, a division of the Simons Foundation. We also acknowledge using the MareNostrum BSC supercomputer of the Spanish Supercomputing Network via project RES/BSC Call AECT-2024-2-0011 (PI AEL). FDS acknowledges support from a Marie Curie Action of the European Union (Grant agreement 101030103). This research has made use of the NASA Exoplanet Archive, which is operated by the California Institute of Technology, under contract with the National Aeronautics and Space Administration under the Exoplanet Exploration Program.

\bibliography{biblio}{}
\bibliographystyle{aasjournal}

\appendix

\section{Rossby number dependence on stellar type}
\label{App: Stellar type dependence}

\begin{figure}[ht!]
	\centering
	\includegraphics[width=.45\hsize]{ 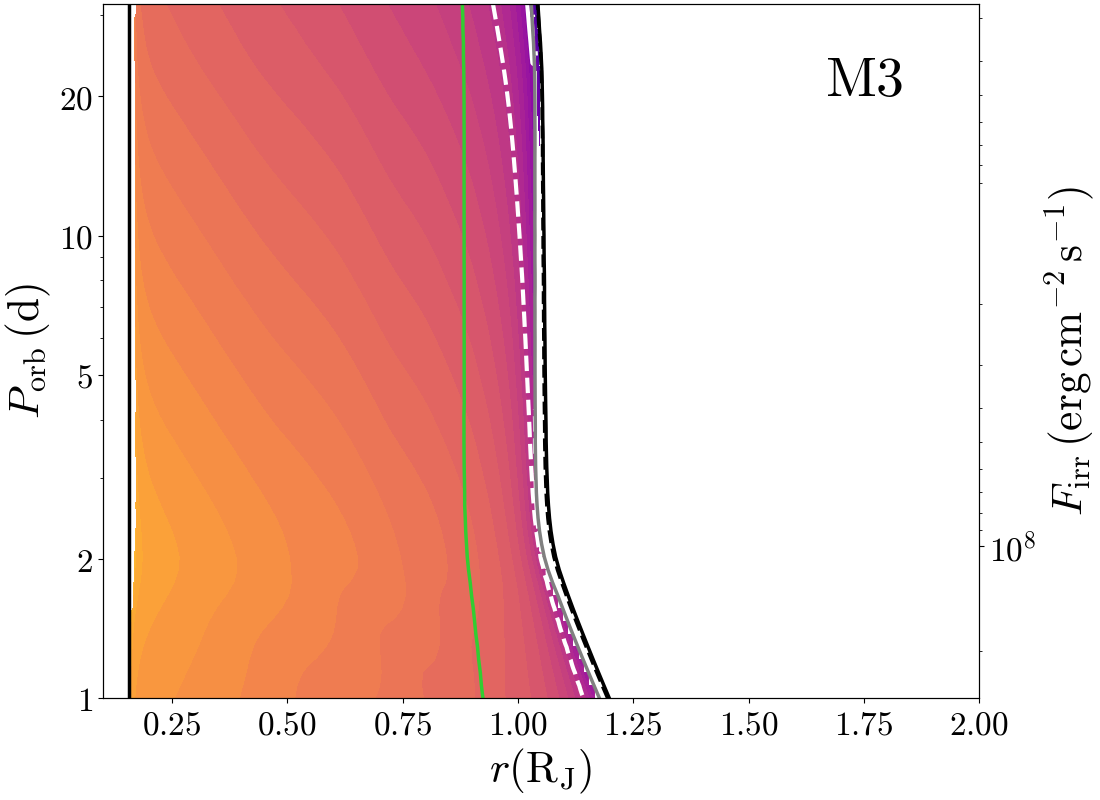}
	\includegraphics[width=.45\hsize]{ 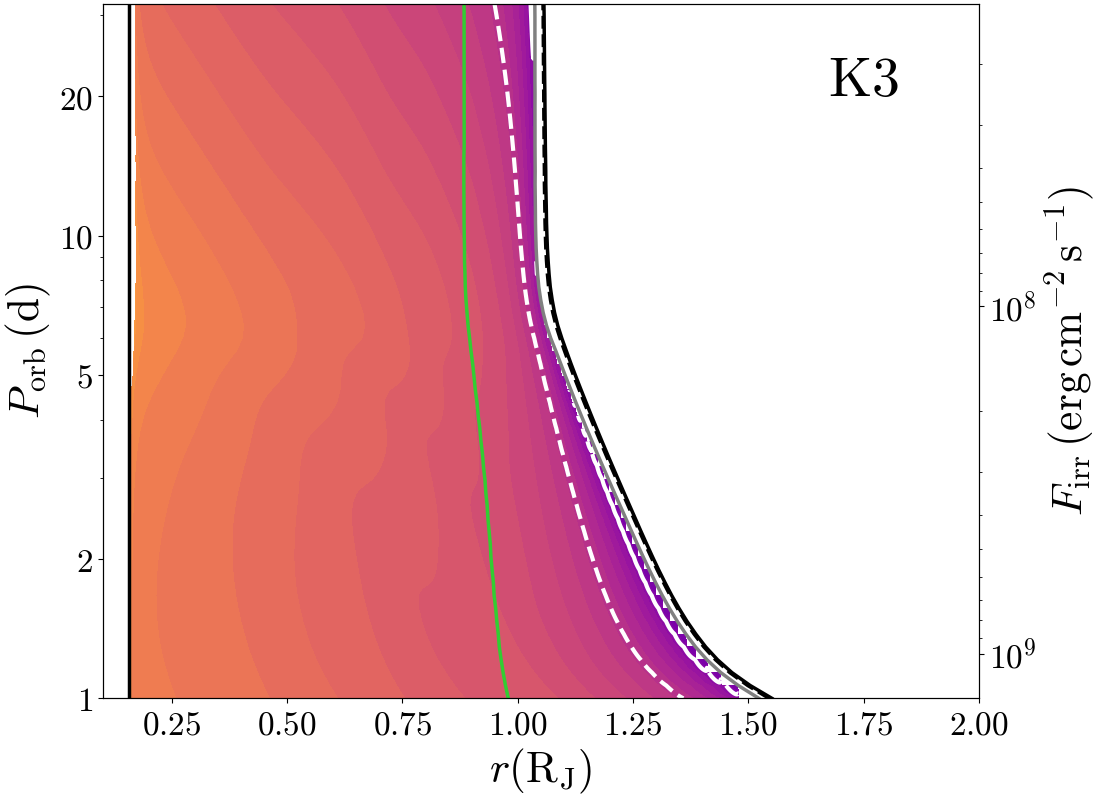}
	\includegraphics[width=.45\hsize]{ 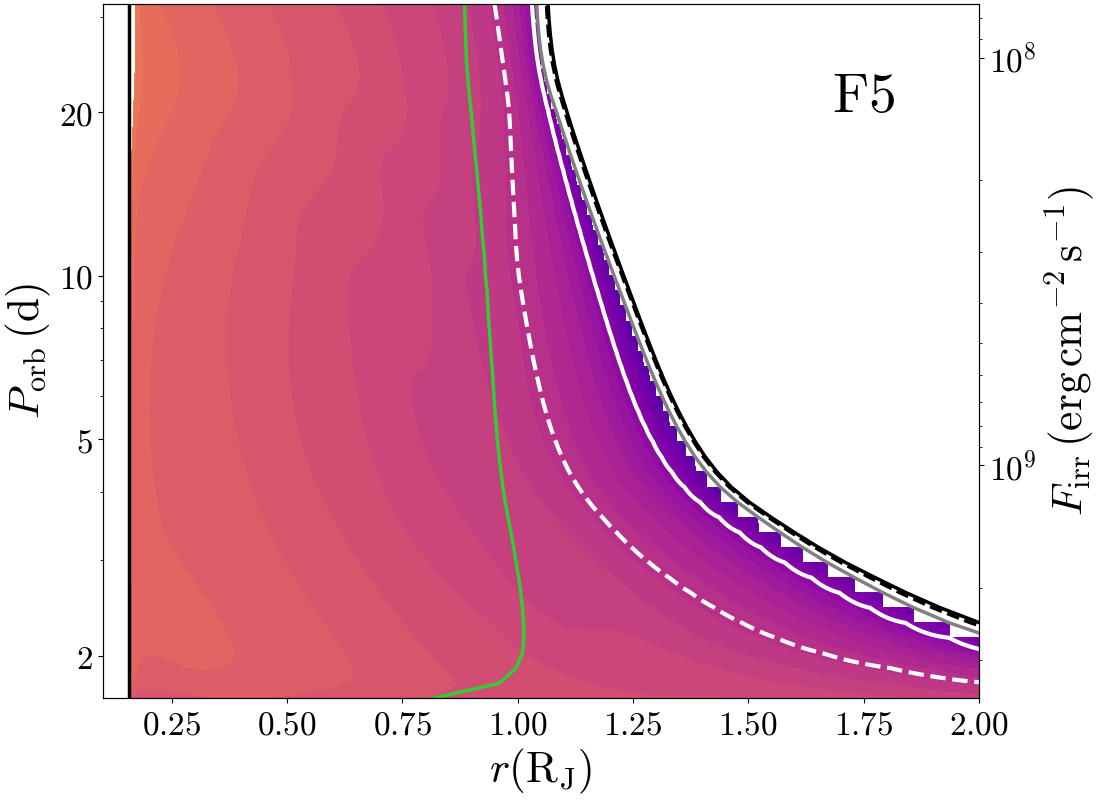}
	\caption{Similar to Fig.~\ref{Fig: Rossby number as a function of distance}: $\mathrm{Ro}$ as a function of $r$ and $a$ for a 1 $\mathrm{M_J}$ planet orbiting the different stars shown in Table~\ref{Tab: HJ common stellar types}. Only the inflated models with heat injection below the irradiated layer are shown.}
	\label{Fig: Rossby number for different stars}
\end{figure}

To evaluate the dependence of $\mathrm{Ro}$ on stellar type, we use the properties of the specific stars shown in Table~\ref{Tab: HJ common stellar types}. Note that, by construction, the evolutionary models are the same but shifted in orbital period, since the relevant parameter is the stellar irradiation, which depends on $P_{\rm orb}$ and stellar mass. The Rossby profiles are instead different, since the $\mathrm{Ro}$ definition includes a further $P_{\rm orb}$ dependency. Similarly to Fig.~\ref{Fig: Rossby number as a function of distance}, in Fig.~\ref{Fig: Rossby number for different stars} we show $\mathrm{Ro}$ as a function of planetary depth and orbital period for an M3, K3, and F5 stars. As expected, the more massive the star, the more inflated the planets are, which leads to larger values of $\mathrm{Ro}$, though they remain smaller than 0.1 for the F5 models.

\end{document}